\documentclass[twocolumn]{aastex62}

\usepackage{amsmath}
\usepackage{amssymb}

\usepackage{lipsum}

\newcommand{\angstrom}{\textup{\AA}}

\def\arcmin{\hbox{$^\prime$}}
\def\arcsec{\hbox{$^{\prime\prime}$}}

\shorttitle{Nebular Spectroscopy of SN\ 2018oh}
\shortauthors{Dimitriadis et al.}

\begin{document}

\title{Nebular Spectroscopy of {\it Kepler}'s Brightest Supernova}

\correspondingauthor{Georgios Dimitriadis}
\email{gdimitri@ucsc.edu}

\author{G.~Dimitriadis}
\affil{Department of Astronomy and Astrophysics, University of California, Santa Cruz, CA 95064, USA}

\author{C.~Rojas-Bravo}
\affil{Department of Astronomy and Astrophysics, University of California, Santa Cruz, CA 95064, USA}

\author{C.~D.~Kilpatrick}
\affil{Department of Astronomy and Astrophysics, University of California, Santa Cruz, CA 95064, USA}

\author{R.~J.~Foley}
\affil{Department of Astronomy and Astrophysics, University of California, Santa Cruz, CA 95064, USA}

\author{A.~L.~Piro}
\affiliation{The Observatories of the Carnegie Institution for Science, 813 Santa Barbara St., Pasadena, CA 91101, USA}

\author{J.~S.~Brown}
\affil{Department of Astronomy and Astrophysics, University of California, Santa Cruz, CA 95064, USA}

\author{P.~Guhathakurta}
\affil{Department of Astronomy and Astrophysics, University of California, Santa Cruz, CA 95064, USA}

\author{A.~C.~N.~Quirk}
\affil{Department of Astronomy and Astrophysics, University of California, Santa Cruz, CA 95064, USA}

\author{A.~Rest}
\affiliation{Space Telescope Science Institute, 3700 San Martin Drive, Baltimore, MD 21218, USA}
\affiliation{Department of Physics and Astronomy, Johns Hopkins University, Baltimore, MD 21218, USA}

\author{G.~M.~Strampelli}
\affiliation{Space Telescope Science Institute, 3700 San Martin Drive, Baltimore, MD 21218, USA}
\affiliation{University of La Laguna,  Calle Padre Herrera, 38200 San Cristóbal de La Laguna, Santa Cruz de Tenerife, Spain}

\author{B.~E.~Tucker}
\affiliation{The Research School of Astronomy and Astrophysics, Mount Stromlo Observatory, Australian National University, Canberra, ACT 2611, Australia}
\affiliation{National Centre for the Public Awareness of Science, Australian National University, Canberra, ACT 2611, Australia}
\affiliation{The ARC Centre of Excellence for All-Sky Astrophysics in 3 Dimensions (ASTRO 3D), Australia}

\author{A.~Villar}
\affiliation{Harvard-Smithsonian Center for Astrophysics, 60 Garden Street, Cambridge, MA 02138, USA}

\begin{abstract}

We present late-time ($\sim$240--260~days after peak brightness) optical photometry and nebular (+236 and +264~days) spectroscopy of SN\ 2018oh, the brightest Type~Ia supernova (SN~Ia) observed by the \textit{Kepler} telescope.  The \textit{Kepler}/K2 30-minute cadence observations started days before explosion and continued past peak brightness. For several days after explosion, SN~2018oh had blue ``excess'' flux in addition to a normal SN rise.  The flux excess can be explained by the interaction between the SN and a Roche-lobe filling non-degenerate companion star.  Such a scenario should also strip material from the companion star, that would emit once the SN ejecta become optically thin, imprinting relatively narrow emission features in its nebular spectrum.  We search our nebular spectra for signs of this interaction, including close examination of wavelengths of hydrogen and helium transitions, finding no significant narrow emission.  We place upper limits on the luminosity of these features of $2.6,\ 2.9\ \mathrm{and}\ 2.1\times10^{37}\ \mathrm{erg\ s^{-1}}$ for H$\alpha$, \ion{He}{1} $\lambda$5875, and \ion{He}{1} $\lambda$6678, respectively.  Assuming a simple models for the amount of swept-up material, we estimate upper mass limits for hydrogen of $5.4\times10^{-4}\ \mathrm{M_{\odot}}$ and helium of $4.7\times10^{-4}\ \mathrm{M_{\odot}}$. Such stringent limits are unexpected for the companion-interaction scenario consistent with the early data.  No known model can explain the excess flux, its blue color, and the lack of late-time narrow emission features.
\end{abstract}

\keywords{supernovae: general ---
supernovae: individual (SN 2018oh)}

\section{Introduction} \label{sec:intro}

The exact nature of the progenitor system for Type Ia supernovae (SNe~Ia) (the ``progenitor problem'')  remains one of the most persistent open questions in stellar evolution. Despite decades of research related to this question, and while SNe~Ia still constitute an extremely powerful probe for measuring the expansion history of the Universe and determine crucial cosmological parameters \citep[e.g.,][]{Riess16ApJ, Jones18, Scolnic18ApJ, DES18}, the stellar systems that lead to the thermonuclear explosion of the carbon/oxygen white dwarf \citep[WD;][]{Hoyle60, Colgate69, Woosley86} and the associated explosion mechanisms are unclear.

In general, two main channels of progenitor systems have been proposed: the single-degenerate (SD) scenario, where the WD explodes due to a thermonuclear runaway near the Chandrasekhar mass ($\mathrm{M_{Ch}}$) by accreting material from a non-degenerate companion \citep[e.g.,][]{Whelan73ApJ}, and the double-degenerate (DD) scenario, where the SN results from the merger of two WDs \citep[e.g.;][]{Iben84ApJS}. Confusing the matter, radiative transfer calculations of explosion models from both scenarios are able to broadly reproduce the basic photometric and spectroscopic properties of SNe~Ia \citep[e.g.][]{Kasen09Natur,Woosley11ApJ,Hillebrandt13FrPhy,Sim13MNRAS}.  We have not yet directly observed the progenitor system of a SN~Ia, and thus we must rely on indirect measures.

\citet{Kasen10ApJ} showed that if the progenitor system contains a non-degenerate, Roche-Lobe filling companion, the SN ejecta will collide with the companion star, and the shock interaction at its surface will produce strong X-ray/UV emission at the first days after the explosion detectable for some viewing angles.  This will result in a luminosity excess beyond the flux expected from the main source of the SN luminosity, $^{56}$Ni radioactive decay. Observationally, this manifests as a two-component rising light curve, with varying component strengths and durations that depend on the size of the companion, the separation of the binary, and the viewing angle.

Additionally for such a scenario, material from the companion's surface will be swept up by the ejecta.  Once the ejecta become optically thin, the companion-star material will emit producing strong, relatively narrow emission features superimposed on an otherwise typical nebular SN~Ia spectrum. Starting with \citet{Marietta00ApJS}, who  were the first to indicate that this emission is anticipated, several theoretical models and simulations have been developed \citep[e.g.,][]{Pan12ApJ, Liu13ApJ, Lundqvist13MNRAS, Botyanszki18ApJ}, predicting emission lines of H$\alpha$, \ion{He}{1} $\lambda\lambda5875$,6678, [\ion{O}{1}] $\lambda\lambda6300$,6364 and/or [\ion{Ca}{2}] $\lambda\lambda7291$,7324, depending on the nature of the companion (whether it is a main-sequence, red-giant or helium star) and the properties of the binary system, with different treatments of the simulations predicting varying strengths and shapes of the emission lines.

These two observational diagnostics have been the subject of numerous studies of early- and late-time SN~Ia observations. Statistical sample studies \citep{Hayden10ApJ1, Ganeshalingam11MNRAS, GonzalezGaitan12, Firth15MNRAS,Olling15Natur} of the early rise times have found slight deviations from the expected $L\propto t^{2}$ law \citep{Arnett82ApJ, Riess99AJ}, attributed to moderate mixing of radioactive $^{56}$Ni into the outer-most layers of the explosion.

Focusing on individual events, SNe 2009ig \citep{Foley12ApJ} and 2011fe \citep{Nugent11, Bloom12} exhibit the expected smomth single-power-law rise of the ligth curve (close to $L\propto t^{2}$) with red early-time colors, providing upper limits on the separation of a potential companion and ruling out evolved stars beyond the giant branch. On the other hand, there are two well-studied SNe~Ia (2012cg and 2017cbv) that show an early blue flux excess.  Those observations can be explained by the interaction of a SN with a 6~$\mathrm{M_{\odot}}$ main-sequence star \citep{Marion16ApJ} and a subgiant companion \citep{Hosseinzadeh17ApJ}, respectively.

Interestingly, \citet{Stritzinger18ApJ} suggest two distinct populations of SNe~Ia, which can be split based on their early ($t < 5$~days after explosion) colors, with the ones with blue early-time colors having brighter peak luminosities and 91T-like spectra and the ones with redder colors having lower peak luminosities and spectra similar to that of typical SNe~Ia. They argue that the interaction scenario cannot produce such a clear dichotomy of peak luminosity for these groups, suggesting that opacity differences in the outer layers of the ejecta, causing faster surface heating, can create distinct colors.

Several different studies have examined the late-time spectra of SNe~Ia, searching for swept-up material from a companion \citep{Mattila05AA, Leonard07ApJ, Shappee13ApJ, Lundqvist13MNRAS, Maguire16MNRAS, Graham17MNRAS, Shappee18ApJ, Sand18ApJ}.  To date, no definitive narrow features have been seen in any of the 18 relatively normal SNe~Ia with late-time spectra, including for SN~2017cbv \citep{Sand18ApJ}, which had excess, blue flux in the few days after explosion \citep{Hosseinzadeh17ApJ}.  The line luminosity limits for these objects correspond to upper limits on the amount of swept-up hydrogen to be $<$1$\times10^{-4}$ -- $5.8\times10^{-2}\ \mathrm{M_{\odot}}$.

SN~2018oh \citep{Dimitriadis18arXiv,Shappee18arXiv,Li18arXiv} is the most recent normal SN Ia showing a blue early rise component. It was discovered by the All Sky Automated Survey for SuperNovae \citep[ASAS-SN,][]{Shappee14ApJ}, with discovery name ASASSN-18bt, and classified as a normal SN~Ia \citep{Leadbeater18TNS, Zhang18ATel} a week before maximum. Its host, UGC~4780, is a small (M = $4.68^{+0.33}_{-0.61}\times10^{8}\ \mathrm{M_{\odot}}$) star-forming (${\rm SFR} \lesssim 0.05\ \mathrm{M_{\odot}\ yr^{-1}}$) galaxy at a redshift of $z = 0.010981$. From the ground-based optical/NIR photometry and spectra, \citet{Li18arXiv} measure a decline rate of $\mathrm{\Delta m_{15}}= 0.96\pm0.03$ mag and a distance modulus of $\mu = 33.61 \pm 0.05$~mag, corresponding to a distance of $52.7 \pm 1.2$~Mpc. SN~2018oh was located in the K2 Campaign 16 field, and its host galaxy was chosen to be monitored by \textit{Kepler} \citep{Haas10ApJ} as part of the K2 Supernova Cosmology Experiment (SCE).

The K2 light curves of SN~2018oh are uniquely informative. The SN is detected within hours after the explosion and is continuously imaged for $\sim$1~month with a 30-min cadence. The most interesting feature observed in the K2 light curve is a prominent two-component rise \citep{Dimitriadis18arXiv,Shappee18arXiv}.  Initially, the flux increased linearly, but after several days, the flux increased quadratically.  Ground-based images \citep{Dotson18RNAAS} show that the SN was particularly blue during the period of the flux excess \citep{Dimitriadis18arXiv}.

In this Letter, we present late-time optical photometry and two nebular spectra of SN~2018oh.  Examining the spectra, we find no narrow emission features indicative of swept-up material and place constraints on the amount of swept-up material in the SN ejecta.

Throughout this paper, we adopt the AB magnitude system, unless where noted, and a Hubble constant of $H_{0} = 73$~km~s$^{-1}$~Mpc$^{-1}$.

\section{Observations and data reduction} \label{sec:observations}

In this section, we present new late-time photometry and spectroscopy of SN\ 2018oh.

\subsection{Late-time Photometry} \label{sec:late_time_phot}

We observed SN~2018oh with the Swope 1.0-m telescope, located at the Las Campanas Observatory, on 2018 Oct 15, Oct 17, and Nov 1 (all times here and later are UT), in $gri$, although not all filters on all dates. Our images were reduced, resampled and calibrated using the \textsc{photpipe} photometric package \citep{Rest05ApJ, Rest14ApJ}, which performs photometry using \textsc{DoPhot} \citep{Schechter93PASP} on difference images. At the time of our observations, the SN was becoming visible after being behind the Sun, and therefore the images were obtained at relatively high airmass (1.98--2.88). Absolute flux calibration was achieved using Pan-STARRS1 \citep[PS1;][]{Chambers16arXiv, Magnier16arXiv, Waters16arXiv} standard stars in the same field as SN~2018oh. In order to remove background contamination from the host galaxy, UGC~4780, we used PS1 $gri$ template images to subtract the host-galaxy emission with {\tt hotpants} \citep{Becker15}.

\begin{figure*}
\begin{center}
  \includegraphics[width=0.95\textwidth]{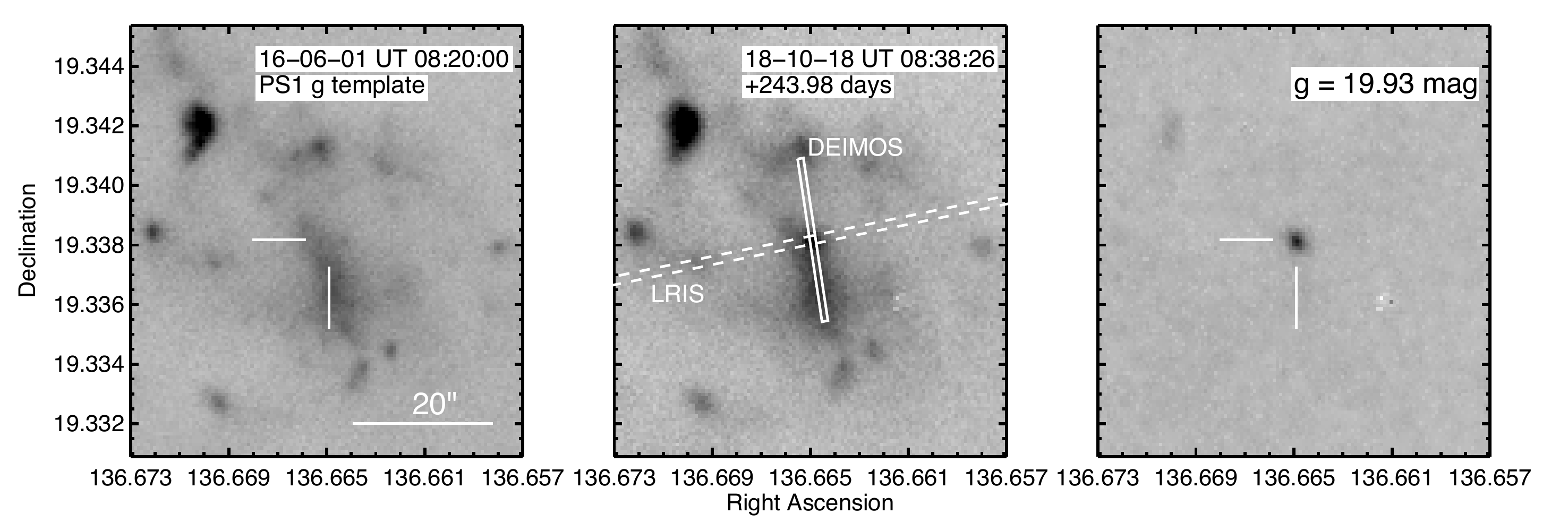}
  \caption{0.87\arcmin\ $\times$ 0.87\arcmin\ PS1 $g$-band template (left), $g$-band Swope image centered on SN~2018oh taken $\sim$244~days after $B$-band maximum brightness (center) and the resulting difference image (right). At this time, SN~2018oh had a $g$-band brightness of $19.93 \pm 0.02$~mag. The solid and dashed white lines boxes represent the slit length, width and orientation for the DEIMOS and LRIS observations, respectively (note that the LRIS slit length is much larger than the image). The position of the SN, at the PS1 template (left) and the difference image (right), is marked with tick marks, while for the science image (center) is at the intersection of the two slits.}
  \label{fig:SN2018oh_images}
\end{center}
\end{figure*}

We show a late-time $g$-band image of SN~2018oh and the PS1 $g$-band template image in Figure~\ref{fig:SN2018oh_images}. Our SN~2018oh photometry is presented in Table~\ref{tab:late_time_photometry}.

\begin{deluxetable}{lccc}[t!]
\tablecaption{SN~2018oh late-time photometry. \label{tab:late_time_photometry}}
\tablecolumns{4}
\tablenum{1}
\tablewidth{0pt}
\tablehead{
\colhead{MJD} &
\colhead{Phase} &
\colhead{Filter} &
\colhead{Brightness} \\
\colhead{} &
\colhead{(Rest-frame days)} &
\colhead{} &
\colhead{(mag)} }
\startdata
58407.37 & +242.00 & $r$ & $21.29 \pm 0.14$ \\
58409.36 & +243.98 & $g$ & $19.93 \pm 0.02$ \\
58424.35 & +258.81 & $r$ & $21.97 \pm 0.16$ \\
58424.36 & +258.82 & $i$ & $21.42 \pm 0.18$ \\
58441.31 & +275.58 & $r$ & $22.12 \pm 0.14$ \\
58441.32 & +275.59 & $i$ & $21.36 \pm 0.10$ \\
58441.33 & +275.60 & $g$ & $20.42 \pm 0.03$ \\
\enddata
\end{deluxetable}

\begin{figure}
\begin{center}
  \includegraphics[width=0.45\textwidth]{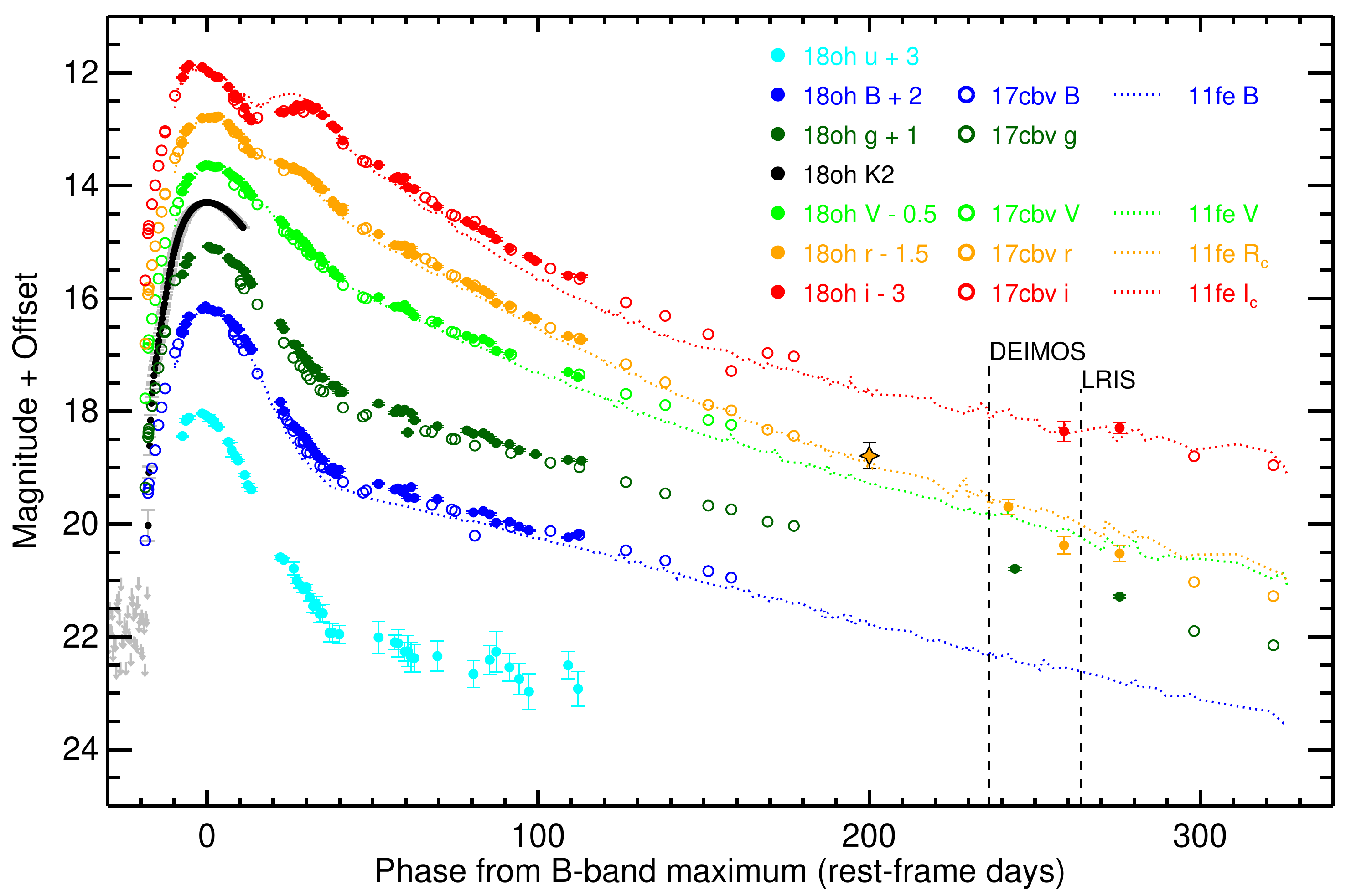}
  \caption{Swope $uBV\!gri$ light curves of SN~2018oh (full circles; offsets noted in the legend) compared to those of SN~2011fe (dotted line) and 2017cbv (open circles), where the light curves of the comparison SNe have been shifted to match the peak of SN~2018oh in each filter. The vertical black dashed lines correspond to the time of our DEIMOS and LRIS spectra, as labeled on the figure. The orange star is our estimate of SN~2018oh's $r$-band magnitude at 200~days after $B$-band maximum brightness (see Section~\ref{sec:sn2018oh_mass_limits}).}
  \label{fig:SN2018oh_light_curves}
\end{center}
\end{figure}

Figure~\ref{fig:SN2018oh_light_curves} displays the complete $uBV\!gri$ Swope light curve of SN\ 2018oh, spanning from $-7.5$ to +110~days relative to peak $B$ brightness (presented in \citet{Li18arXiv}), with the addition of the new late-time data presented here. The light curves have been corrected for Milky Way extinction using the \citet{Fitzpatrick99PASP} law (with $R_{V} = 3.1$) for $E(B-V)_{MW} = 0.037$~mag and placed in rest-frame using $z = 0.011$. In a similar manner to \citet{Dimitriadis18arXiv}, we compare the light curves of SN~2018oh to those of SNe~2011fe \citep[BV$R_{C}I_{C}$;][]{Munari13NewA} and 2017cbv ($BV\!gri$; Rojas-Bravo et al., in prep.).  These SNe represent a typical SN~Ia with no initial flux excess and a SN~Ia with a prominent blue flux excess, respectively.  Despite the differences in the first few days after explosion, all three SNe behave similarly, from peak brightness until the epoch of our latest SN~2018oh data.

\subsection{Late-time Spectroscopy} \label{sec:late_time_spec}

We obtained two optical spectra of SN~2018oh: one with the DEep Imaging Multi-Object Spectrograph \citep[DEIMOS;][]{Faber03SPIE} and one with the Low Resolution Imaging Spectrometer \citep[LRIS;][]{Oke95PASP}, mounted on the 10-meter Keck~II and Keck~I telescopes at the W.\ M.\ Keck Observatory, respectively. The DEIMOS spectrum consists of two 30-minute exposures, taken on 2018 Oct 10 and 11 (at an average phase of +236.2 days after $B$-band maximum brightness) and covers a (4620 --- 9830~\AA) wavelength range. We used an 0.8\arcsec-wide slitlet with the 600ZD grating (central wavelength of 7200~\AA) and the GG455 order-blocking filter, with the exposure being taken on the paralactic angle. The data were reduced using a modified version of the DEEP2 data-reduction pipeline \citep{Cooper12ascl, Newman13ApJS}, which bias-corrects, flattens, rectifies, and sky-subtracts the data. The LRIS spectrum consists of one 60-minute exposure, taken on 2018 Nov 19 (at an average phase of +264.0 days after $B$-band maximum brightness), and covers a (3300 --- 10,100~\AA) wavelength range. We used the 1.0\arcsec-wide slit, the 600/4000 grism (blue side), the 400/8500 grating (red side, central wavelength at 7743 \AA) and the D560 dichroic. For that exposure, we used an angle of 170 degrees (north to east), in order to minimize the host-galaxy light contribution, benefiting from the Atmospheric Dispersion Compensator (ADC) module of Keck I, that allows LRIS to operate with reduced differential refraction. These data were reduced using standard \textsc{iraf}\footnote{IRAF is distributed by the National Optical Astronomy Observatory, which is operated by the Association of Universities for Research in Astronomy (AURA) under a cooperative agreement with the National Science Foundation.} for bias corrections and flat fielding. For both of the spectra, we employed our own IDL routines to flux calibrate the data and remove telluric lines using the well-exposed continua of the spectrophotometric standards \citep[e.g.,][]{Foley03PASP, Silverman12MNRAS}. 

\begin{figure}
\begin{center}
  \includegraphics[width=0.45\textwidth]{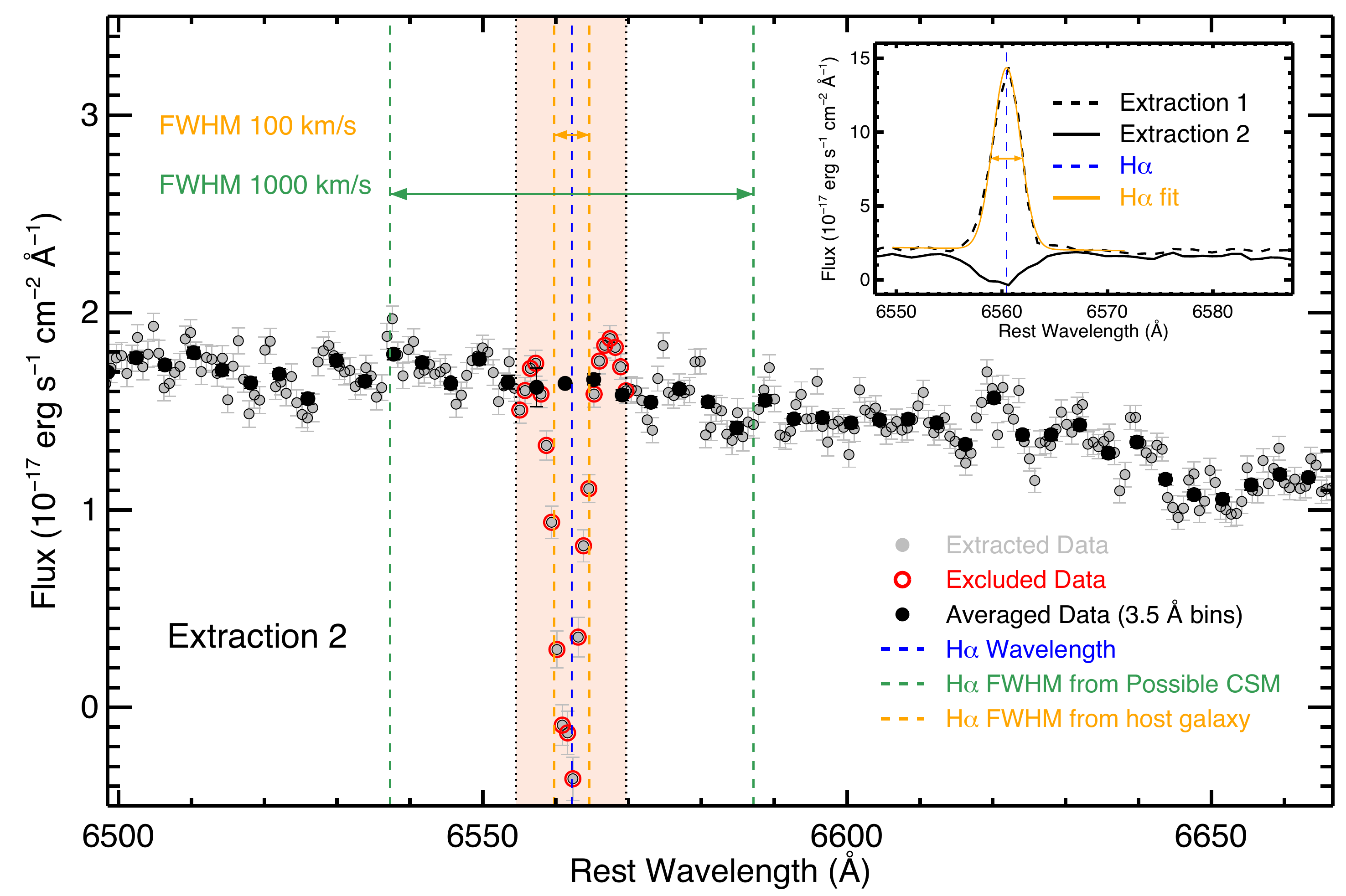}
  \caption{DEIMOS spectrum of SN~2018oh (grey circles) near the position of H$\alpha$.  This is our first extracted spectrum where the emission line is oversubtracted.  This spectrum and the spectrum resulting from the more distant background regions are shown in the inset (in black and gold, respectively).  We fit a Gaussian to the emission feature and display its FWHM as vertical gold-dashed lines.  The FWHM of this line is significantly smaller than of a feature expected from swept-up material ($\sim$1000 $\mathrm{km \ s^{-1}}$; indicated by the vertical green-dashed lines).  Data within $3\times{\rm FWHM}$ (marked by the vertical black-dotted lines and the peach background) are outlined in red and are removed from our final spectrum.  The full black circles correspond to the spectrum with 3.5-\AA\ binning, which we use in Section~\ref{sec:sn2018oh_mass_limits}.}
  \label{fig:SN2018oh_spec_clip}
\end{center}
\end{figure}

Because the SN is embedded in diffuse galactic light, we had to carefully extract the spectra to mitigate host-galaxy contamination.  To do this, we extracted the SN spectrum using two sets of background regions.  One has the background regions close to the SN position, which provides an excellent representation of the continuum flux at the SN position.  However, since these regions have stronger emission flux than at the SN position, strong emission lines are oversubtracted.  To compensate for this effect, we also extracted the SN spectrum with the background regions further from the SN position, which does not fully remove the galactic light, but also provides an accurate measurement of the emission flux at the SN position. Using the second extraction that has undersubtracted galactic emission features, we fit a Gaussian to the H$\alpha$ emission line (Figure~\ref{fig:SN2018oh_spec_clip}), finding a FWHM of $100 \pm 20\ \mathrm{km \ s^{-1}}$. This line is significantly narrower than what is predicted for the interaction scenario ($\sim$1000 $\mathrm{km \ s^{-1}}$), indicating that it originates from the galaxy.  We then replace all data within $3\times{\rm FWHM}$ in the first extraction (that originally had oversubtracted features) using a linear fit to the remaining dataset, and finally rebinning the spectrum to 3.5 \AA, to obtain our final spectra.

\begin{figure*}
\begin{center}
  \includegraphics[width=0.9\textwidth]{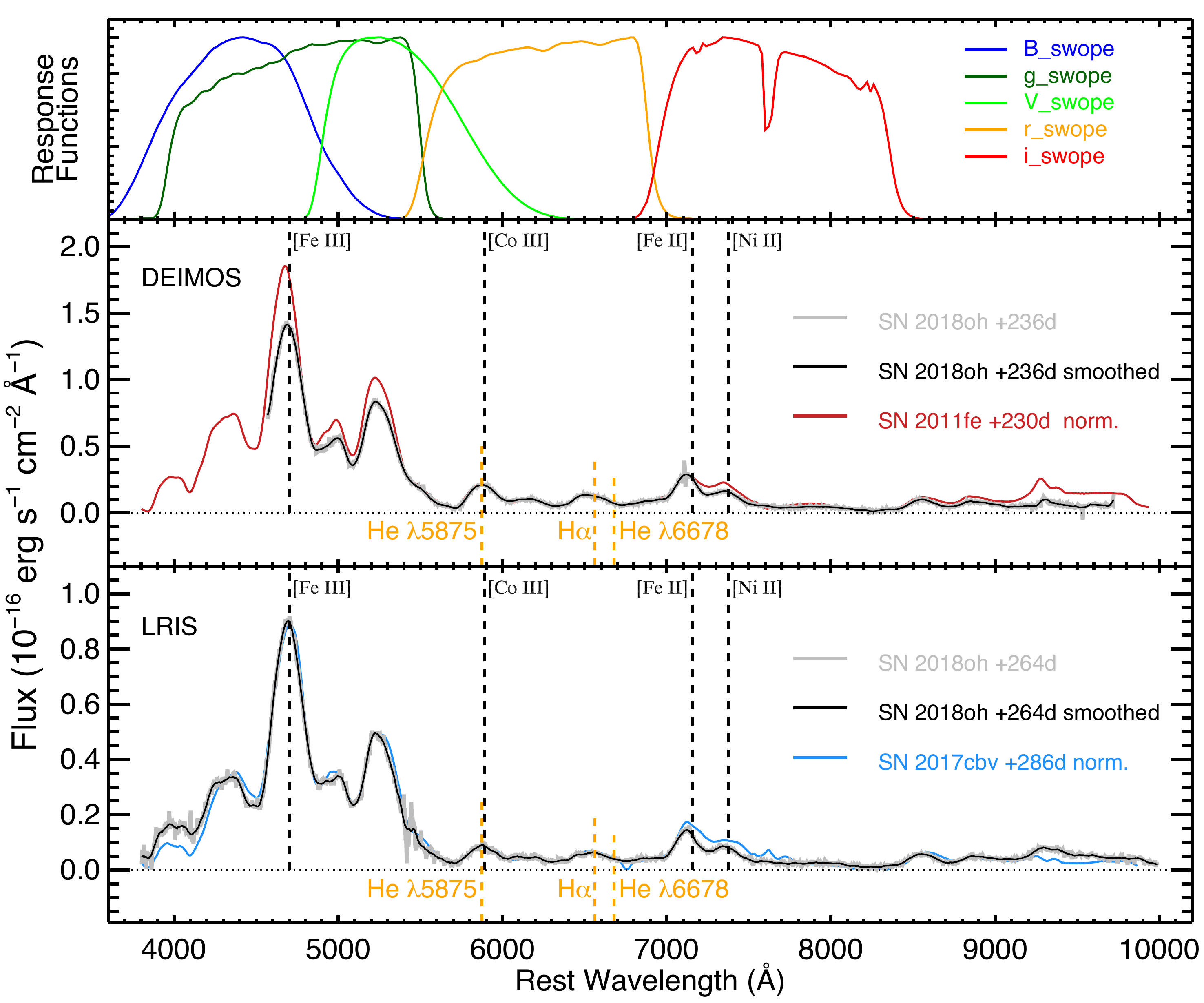}
  \caption{Rest-frame nebular-phase (+236, middle panel, and 264, bottom panel, days form B-band maximum) spectra of SN\ 2018oh. We show the 3.5\AA-binned spectra with grey and the 100\AA-smoothed with black solid lines, respectively. We compare the spectra with the +230 days SN\ 2011fe spectrum \citep[red;][]{Graham15MNRAS} and the +286 days SN\ 2017cbv spectrum \citep[blue;][]{Sand18ApJ}. All spectra are normalized to the flux of SN\ 2018oh in the $r$-band of the Swope telescope at its corresponding phase, for which we plot the response functions of its broad-band filters at the top panel. We also mark four iron-peak elements' nebular-phase lines (vertical dashed black lines) and the three zero-velocity positions of the expected interaction lines from Hydrogen and Helium in the interaction scenario (vertical dashed orange lines).}
  \label{fig:SN2018oh_spectra}
\end{center}
\end{figure*}

The nebular spectra of SN~2018oh are shown as the solid grey lines in Figure~\ref{fig:SN2018oh_spectra}.  The spectra have been corrected for Milky-Way reddening with the same \citet{Fitzpatrick99PASP} law as our photometry, and smoothed using a second-order, 100-\AA-wide Savitzky–Golay smoothing polynomial, shown with a solid black line.  We additionally overplot two late-time spectra of SNe~2011fe \citep{Graham15MNRAS} and 2017cbv \citep{Sand18ApJ} at +230 and +286~days respectively, corrected and smoothed in the same manner and scaled to the $r$-band flux of SN~2018oh.

We do not detect any relatively narrow hydrogen or helium emission features originating from swept-up material.  We determine the flux limits for these features and mass limits for the material below.

\section{Mass Limits For Swept-up Material} \label{sec:sn2018oh_mass_limits}

In order to provide statistical constraints on the amount of stripped material from a potential non-degenerate companion, we follow the methodology of \citet{Sand18ApJ}. This approach is similar to previous works \citep{Shappee13ApJ, Maguire16MNRAS, Graham17MNRAS, Shappee18ApJ}, but uses recent multi-dimensional radiative transfer models and hydrodynamical simulations of ejecta-companion interaction from \citet{Botyanszki18ApJ} instead of simpler treatments based on the models of \citet{Mattila05AA} and \citet{Lundqvist13MNRAS}. \citet{Botyanszki18ApJ} uses the \citet{Boehner17MNRAS} hydrodynamical models of a SN~Ia interacting with its companion and synthesizes the resulting spectra at +200~days after peak. 

\begin{figure}
\begin{center}
  \includegraphics[width=0.45\textwidth]{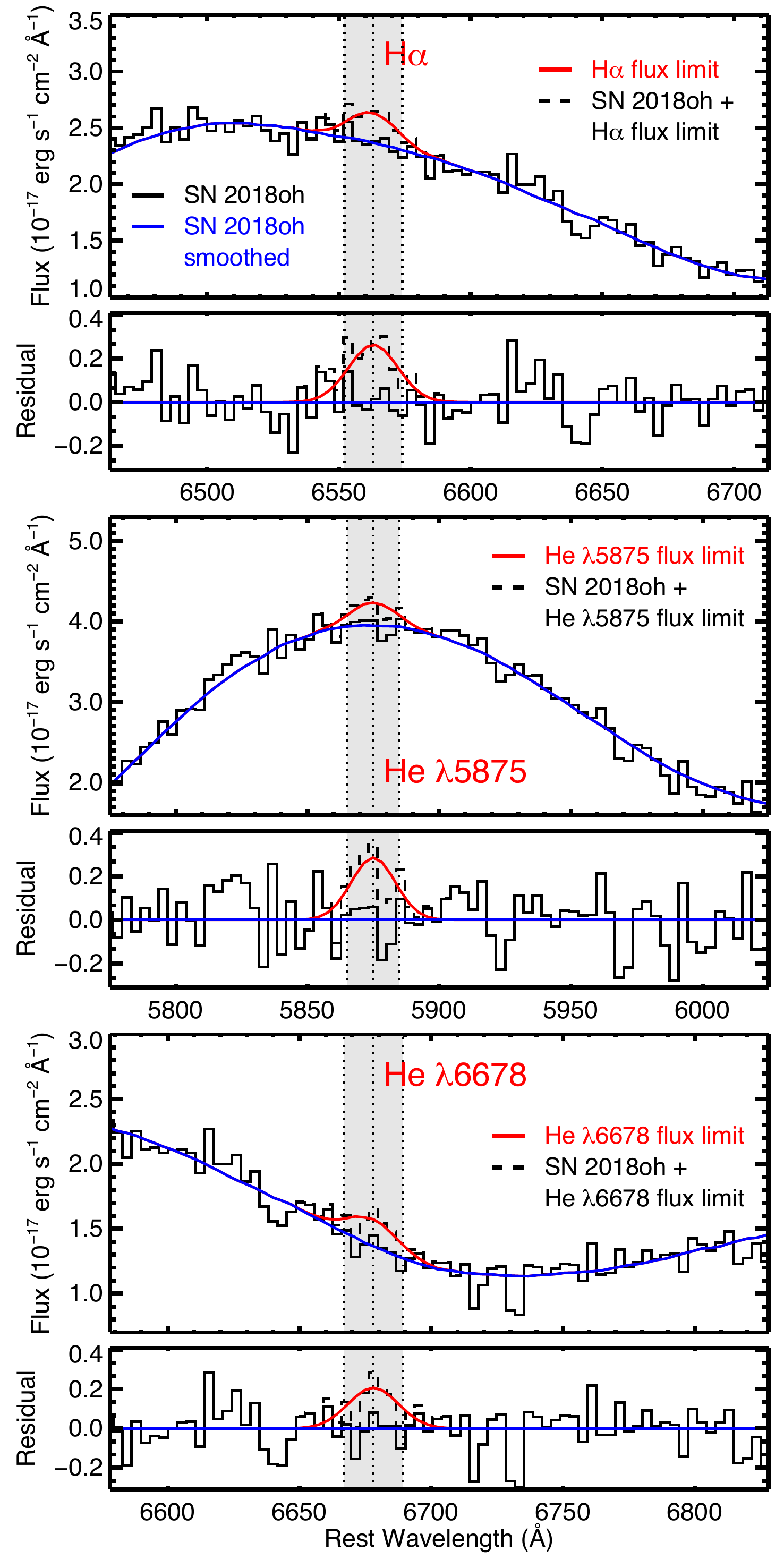}
  \caption{The DEIMOS spectrum of SN~2018oh zoomed in to show H$\alpha$ (top panel), \ion{He}{1} $\lambda5875$ (middle panel), and \ion{He}{1} $\lambda6678$ (bottom panel).  The flux is adjusted to what would have been seen 200~days after maximum brightness. The underlying continuum, which we approximate with a Savitsky–-Golay smoothed (with a 180-\AA\ filter) version of the spectrum, is displayed as blue solid lines. The grey-shaded region corresponds the $\pm22$\AA\ (roughly 1000~$\mathrm{km \ s^{-1}}$) region around the rest wavelength of each line.  In red, we insert a feature at the rest-wavelength of each line with a FWHM of 1000~$\mathrm{km \ s^{-1}}$ and a height corresponding to our 3-$\sigma$ detection limit above the smoothed continuum.  With a dashed black line, we also show how such a feature would look in the data.  For each feature, we also display the residuals relative to the continuum.}
  \label{fig:SN2018oh_hydrogen_helium_limit}
\end{center}
\end{figure}

Despite the \citet{Botyanszki18ApJ} spectrum being generated for an epoch of 200~days after peak and our spectrum being from 230~days after peak, we can still easily compare the data to the models.  Since SN~Ia spectral features do not change significantly between these two epochs, we can assume that SN~2018oh had the same spectral shape at +200~days as it has in our spectrum.  To appropriately scale the flux, we simply interpolate our $r$-band Swope light curve (which wavelength range covers the hydrogen and helium lines we are interested in) to determine the brightness at +200~days, finding $r_{200\mathrm{d}} = 20.40 \pm 0.23$ mag. Finally, we bin our spectra to 3.5~\AA, similar to \citet{Sand18ApJ}, so that we can directly compare SN~2018oh to SN~2017cbv. The final DEIMOS spectrum is shown as a black solid line in Figure~\ref{fig:SN2018oh_hydrogen_helium_limit}.

As seen in Figure~\ref{fig:SN2018oh_spectra}, the hydrogen and helium emission lines coincide with broad emission features from the SN.  Thus, in order to appropriately determine the continuum of this underlying emission, we use a second-order Savitsky--Golay smoothing polynomial filter with a window of 180 \AA.  We repeated our analysis with varying window widths (80 to 260 \AA), and our final mass estimates are well within 1-$\sigma$ of our initial choice. We will continue our analysis with the 180 \AA~window width to ease comparison with the \citet{Sand18ApJ} study of SN~2017cbv.

Examining the unsmoothed DEIMOS spectrum, both in isolation and compared to the smoothed spectrum, we detect no obvious emission features expected from the interaction scenario.  To determine the flux limit for these features, we first measure the RMS noise in the residual (data-continuum) spectra. We approximate these emissions as Gaussians with a peak flux of 3$\times$RMS, at the spectral region of H$\alpha$, \ion{He}{1} $\lambda5875$, and \ion{He}{1} $\lambda6678$ and a FWHM of 22~\AA\ (corresponding to $1000\ \mathrm{km\ s^{-1}}$).  Our nominal 3-$\sigma$ flux limit for H$\alpha$, \ion{He}{1} $\lambda5875$, and \ion{He}{1} $\lambda6678$ are $2.6,\ 2.9\ \mathrm{and}\ 2.1\times10^{-18}\ \mathrm{erg\ s^{-1}\ cm^{-2} \angstrom^{-1}}$, respectively.  Adopting the luminosity distance computed in \citet{Li18arXiv} the luminosity limits are $2.1,\ 2.0\ \mathrm{and}\ 1.6\times10^{37}\ \mathrm{erg\ s^{-1}}$, respectively, and converting the luminosity limits to mass limits using Equation~1 of \citet{Botyanszki18ApJ}, we determine SN~2018oh had maximum stripped hydrogen and helium masses of $5.4\times10^{-4}\ \mathrm{M_{\odot}}$ and $4.7\times10^{-4}\ \mathrm{M_{\odot}}$, respectively. By adopting 1-$\sigma$ uncertainties of the SN brightness at 200 days and of the luminosity distance, we estimate flux limits of $3.2,\ 3.5\ \mathrm{and}\ 2.5\times10^{-18}\ \mathrm{erg\ s^{-1}\ cm^{-2} \angstrom^{-1}}$, luminosity limits of $2.6,\ 2.5\ \mathrm{and}\ 2.1\times10^{37}\ \mathrm{erg\ s^{-1}}$ and mass limits of $6.4\times10^{-4}\ \mathrm{M_{\odot}}$ and $5.5\times10^{-4}\ \mathrm{M_{\odot}}$. We repeated our analysis with our LRIS spectrum, taken at +264 days from maximum, deriving similar mass limits ($M_{H}<6.5\times10^{-4}\ \mathrm{M_{\odot}}$ and $M_{He}<8.4\times10^{-4}\ \mathrm{M_{\odot}}$), thus we continue our analysis with the DEIMOS +232 days spectrum.

We additionally provide the mass limit of hydrogen, derived using the method of \citet{Leonard07ApJ}, for which the authors use the models from \citet{Mattila05AA}. \citet{Mattila05AA} estimate that, at +380 days from peak brightness, a gaussian emission line of $3.36\times10^{35}\ \mathrm{erg\ s^{-1}\ \angstrom^{-1}}$ is expected from 0.05 $\mathrm{M_{\odot}}$ of stripped hydrogen. By scaling our DEIMOS spectrum to that epoch, adopting the linear decline rate of a factor of 4 at 200--300 days, we derive an equivalent width of that feature of $W_{\lambda}(0.05\ \mathrm{M_{\odot}})=25.53$ \AA, while the equivalent width of the strongest gaussian emission line of our spectrum that could remain undetected, at that region, is $W_{\lambda}(3\sigma)=0.99$ \AA. Finally, adopting the linear scale between the mass of hydrogen and the equivalent width of the emission line, we derive the upper mass limit $M_{H}<1.9\times10^{-3}\ \mathrm{M_{\odot}}$.

\section{Discussion and conclusions} \label{sec:discussion_conclusions}

We have presented late-time photometry and spectroscopy of the closest SN observed by \textit{Kepler}, SN\ 2018oh, which exhibits a prominent early linearly rising light curve, before settling back to a typical $L \propto t^{2}$ rise. Examining the spectrum, we do not detect the relatively narrow emission expected when a SN interacts with a close, non-degenerate companion and sweeps up material from the companion's outer layers.  After flux-calibrating our nebular spectra to Swope photometry, assuming that the companion star is Roche-lobe filling, and using the models of \citep{Botyanszki18ApJ}, we determine 3-$\sigma$ upper limits for the mass of swept-up hydrogen and helium of $5.4\times10^{-4}\ \mathrm{M_{\odot}}$ and $4.7\times10^{-4}\ \mathrm{M_{\odot}}$, respectively.

\citet{Dimitriadis18arXiv} consider two possible physical mechanisms that adequately reproduce the early \textit{Kepler}/K2 light curve: interaction with a companion at a distance of $a = 2\times10^{12}$~cm and $0.03\,$M$_{\odot}$ of $^{56}$Ni in the outer layers of the ejecta.  While both of these mechanisms were considered possible, the surface $^{56}$Ni model cannot easily reproduce the blue color observed in the first few days.  Because of the color constraint, \citet{Dimitriadis18arXiv} slightly favored the interaction scenario.

Assuming the Roche-Lobe filling criterion, \citet{Dimitriadis18arXiv} suggests a subgiant companion with $M \approx 1$--6~M$_\odot$ and $R \approx 10$--15~R$_\odot$. \citet{Botyanszki18ApJ}, using \citet{Boehner17MNRAS} models, provide H$\alpha$ luminosities for various companion stars, with the \citet{Dimitriadis18arXiv} proposed companion star having properties intermediate to models MS38, SG, and RG319. These models predict $L_{H_{\alpha}} = 6.8$, 5.6 and $4.5 \times 10^{39}\ \mathrm{erg\ s^{-1}}$ with $M_{H_{\alpha}} = 0.25$, 0.17, and $0.28\ \mathrm{M_{\odot}}$ respectively.  For SN~2018oh, the H$\alpha$ luminosity is constrained to be two orders of magnitude less than the models.  However, we note that our inferred hydrogen limits are based on the extrapolation of the simulations. Moreover, simulations that cover a wider range of the SD scenario parameter space, such as the binary separation and the companion mass are still lacking.

Our inferred mass limits are in accordance with the recent study of \citet{Tucker18arXiv}, where the authors, analyzing a nebular (+265 rest-frame days after maximum) spectrum of SN~2018oh, estimate  $M_{H}<6\times10^{-3}\ \mathrm{M_{\odot}}$ and $M_{He}<2\times10^{-2}\ \mathrm{M_{\odot}}$.

To date, well--studied SNe~Ia with prominent linear-rise components in their early light curves and particularly early blue colors are SNe~2013dy \citep{Zheng13ApJ}, ASASSN-14lp \citep{Shappee16ApJ}, iPTF16abc \citep{Miller18ApJ}, 2017cbv \citep{Hosseinzadeh17ApJ} and 2018oh \citep{Dimitriadis18arXiv}.  However, no SN in this sample has nebular spectra indicative of companion interaction \citep{Shappee18ApJ, Sand18ApJ}. There are four possible explanations for the combination of early blue excess flux and a lack of strong, relatively narrow hydrogen and helium emission features in the nebular spectra:

\begin{enumerate}  
\item SN~2018oh did not have a Roche-Lobe filling companion.   However, some SNe~Ia clearly have relatively dense CSM as seen by time-variable absorption features (e.g., \citealt{Patat07Sci}; \citealt{Simon09ApJ} and a relatively large fraction of SNe~Ia must be ``gas rich;'' e.g., \citealt{Sternberg11Sci}; \citealt{Foley12ApJ2}; \citealt{Maguire13MNRAS}), yet those SNe also do not have narrow emission features in their nebular spectra.  Furthermore, SNe~Iax \citep{Foley13ApJ} have strong evidence for Roche-lobe filling companions \citep[e.g.,][]{McCully14Natur}, but none have strong hydrogen or helium emission lines in their late-time spectra (\citealt{Foley16MNRAS}; Jacobson-Gal\'an et al., in prep.).  There are also some SNe~Ia that have strong emission lines from circumstellar interaction, including at early times \citep[the ``SN~Ia-CSM'' class; e.g.,][]{Dilday12Sci, Silverman13ApJS}, but this emission is exclusively very strong indicating very dense CSM.  While SN~2018oh may lack a Roche-lobe filling companion, that alone does not explain the lack of hydrogen/helium emission features in other nebular spectra and the lack of weak interaction signatures for some SNe~Ia-CSM. 
\item The current theoretical models of the Roche-Lobe filling SD scenario overpredict the H$\alpha$ luminosity at the times of our data. While these theoretical models cannot fully capture the complex physics involved (asymmetries in the explosion, precise atomic line data, reliable radiative transport codes), it is unlikely that the amount of stripped material predicted is off by two orders of magnitude.  At face value, this explanation seems unlikely.
\item SN~2018oh had a more distant non-degenerate companion (i.e., a symbiotic progenitor system).  Having a more distant companion would reduce the amount of material stripped from its surface.  However, one would need a very unlikely orientation to possibly reproduce the early flux.
\item SN~2018oh had a significant amount of $^{56}$Ni on its surface (to produce the fast rise of the light curve) and radiative transfer calculations incorrectly predict that this light should be red (because of line blanketing from the high abundance of Fe-group elements).  Again, simple calculations show that the excess flux produced in this scenario should be red, inconsistent with SN~2018oh.  Red flux excesses have been seen for other SNe \citep{Jiang17Natur}, further indicating that this basic scenario is correct for at least some events.  An asymmetric distribution of $^{56}$Ni in the outermost layers combined with a particular viewing angle may resolve this issue.
\item Some models are able to reproduce the general properties of SN~2018oh, such as a detached system consisting of a WD and a RG-like companion under the common-envelope wind SD scenario \citep{Meng18ApJ,Meng18arXiv}, or a non-violent DD scenario involving the collision of the SN ejecta with circumstellar material originating from an accretion disk formed during the merger process of the two WDs \citep{Levanon17MNRAS}. However, more detailed modeling of these potentially rare channels, alongside studies involving their rates, is necessary.
\end{enumerate}

Considering several possibilities, we conclude that there are no known models that can simultaneously explain the blue early-time flux excess and the lack of late-time narrow emission lines.  As the population of these remarkable events grows, we will be able to statistically investigate their properties which may reveal other possible explanations \citep[e.g. see][]{Stritzinger18ApJ}. In addition to new discoveries and observations, more realistic theoretical models, with better radiative transfer calculations, are needed.  We will continue observing SN~2018oh and, at the same time, actively pursue to discover other SNe~Ia within hours of explosion, focusing on their early color evolution and spectral evolution from the first few hours to several months after peak brightness. 

\facility{Swope, Keck:I (LRIS), Keck:II (DEIMOS)}

\begin{acknowledgments} 

\bigskip

We thank the anonymous referee for helpful comments that improved the clarity and presentation of this paper.

Some of the data presented herein were obtained at the W.\ M.\ Keck Observatory, which is operated as a scientific partnership among the California Institute of Technology, the University of California and the National Aeronautics and Space Administration. The Observatory was made possible by the generous financial support of the W.\ M.\ Keck Foundation.
The authors wish to recognize and acknowledge the very significant cultural role and reverence that the summit of Maunakea has always had within the indigenous Hawaiian community.  We are most fortunate to have the opportunity to conduct observations from this mountain.

This paper includes data gathered with the 1.0-m Swope Telescope located at Las Campanas Observatory, Chile. We thank J. Anais, A. Campillay and N. M. Elgueta for assistance with these observations.

The UCSC team is supported in part by NASA grants 14-WPS14-0048, NNG16PJ34G, and NNG17PX03C; NSF grants AST-1518052 and AST-1815935; the Gordon \& Betty Moore Foundation; the Heising-Simons Foundation; and by a fellowship from the David and Lucile Packard Foundation to R.J.F.

\end{acknowledgments}

\bibliographystyle{aasjournal} 
\bibliography{SN2018oh_nebular}

\begin{thebibliography}{}
\expandafter\ifx\csname natexlab\endcsname\relax\def\natexlab#1{#1}\fi
\providecommand{\url}[1]{\href{#1}{#1}}

\bibitem[{{Arnett}(1982)}]{Arnett82ApJ}
{Arnett}, W.~D. 1982, \apj, 253, 785

\bibitem[{{Becker}(2015)}]{Becker15}
{Becker}, A. 2015, {HOTPANTS: High Order Transform of PSF ANd Template
  Subtraction}, Astrophysics Source Code Library, , , ascl:1504.004

\bibitem[{{Bloom} {et~al.}(2012){Bloom}, {Kasen}, {Shen}, {Nugent}, {Butler},
  {Graham}, {Howell}, {Kolb}, {Holmes}, {Haswell}, {Burwitz}, {Rodriguez}, \&
  {Sullivan}}]{Bloom12}
{Bloom}, J.~S., {Kasen}, D., {Shen}, K.~J., {et~al.} 2012, \apjl, 744, L17

\bibitem[{{Boehner} {et~al.}(2017){Boehner}, {Plewa}, \&
  {Langer}}]{Boehner17MNRAS}
{Boehner}, P., {Plewa}, T., \& {Langer}, N. 2017, \mnras, 465, 2060

\bibitem[{{Boty{\'a}nszki} {et~al.}(2018){Boty{\'a}nszki}, {Kasen}, \&
  {Plewa}}]{Botyanszki18ApJ}
{Boty{\'a}nszki}, J., {Kasen}, D., \& {Plewa}, T. 2018, \apjl, 852, L6

\bibitem[{{Chambers} {et~al.}(2016){Chambers}, {Magnier}, {Metcalfe},
  {Flewelling}, {Huber}, {Waters}, {Denneau}, {Draper}, {Farrow}, {Finkbeiner},
  {Holmberg}, {Koppenhoefer}, {Price}, {Saglia}, {Schlafly}, {Smartt},
  {Sweeney}, {Wainscoat}, {Burgett}, {Grav}, {Heasley}, {Hodapp}, {Jedicke},
  {Kaiser}, {Kudritzki}, {Luppino}, {Lupton}, {Monet}, {Morgan}, {Onaka},
  {Stubbs}, {Tonry}, {Banados}, {Bell}, {Bender}, {Bernard}, {Botticella},
  {Casertano}, {Chastel}, {Chen}, {Chen}, {Cole}, {Deacon}, {Frenk},
  {Fitzsimmons}, {Gezari}, {Goessl}, {Goggia}, {Goldman}, {Grebel}, {Hambly},
  {Hasinger}, {Heavens}, {Heckman}, {Henderson}, {Henning}, {Holman}, {Hopp},
  {Ip}, {Isani}, {Keyes}, {Koekemoer}, {Kotak}, {Long}, {Lucey}, {Liu},
  {Martin}, {McLean}, {Morganson}, {Murphy}, {Nieto-Santisteban}, {Norberg},
  {Peacock}, {Pier}, {Postman}, {Primak}, {Rae}, {Rest}, {Riess}, {Riffeser},
  {Rix}, {Roser}, {Schilbach}, {Schultz}, {Scolnic}, {Szalay}, {Seitz},
  {Shiao}, {Small}, {Smith}, {Soderblom}, {Taylor}, {Thakar}, {Thiel},
  {Thilker}, {Urata}, {Valenti}, {Walter}, {Watters}, {Werner}, {White},
  {Wood-Vasey}, \& {Wyse}}]{Chambers16arXiv}
{Chambers}, K.~C., {Magnier}, E.~A., {Metcalfe}, N., {et~al.} 2016, ArXiv
  e-prints, arXiv:1612.05560

\bibitem[{{Colgate} \& {McKee}(1969)}]{Colgate69}
{Colgate}, S.~A., \& {McKee}, C. 1969, \apj, 157, 623

\bibitem[{{Cooper} {et~al.}(2012){Cooper}, {Newman}, {Davis}, {Finkbeiner}, \&
  {Gerke}}]{Cooper12ascl}
{Cooper}, M.~C., {Newman}, J.~A., {Davis}, M., {Finkbeiner}, D.~P., \& {Gerke},
  B.~F. 2012, {spec2d: DEEP2 DEIMOS Spectral Pipeline}, Astrophysics Source
  Code Library, , , ascl:1203.003

\bibitem[{{DES Collaboration} {et~al.}(2018){DES Collaboration}, {Abbott},
  {Allam}, {Andersen}, {Angus}, {Asorey}, {Avelino}, {Avila}, {Bassett},
  {Bechtol}, {Bernstein}, {Bertin}, {Brooks}, {Brout}, {Brown}, {Burke},
  {Calcino}, {Carnero Rosell}, {Carollo}, {Carrasco Kind}, {Carretero},
  {Casas}, {Castander}, {Cawthon}, {Challis}, {Childress}, {Clocchiatti},
  {Cunha}, {D'Andrea}, {da Costa}, {Davis}, {Davis}, {De Vicente}, {DePoy},
  {Desai}, {Diehl}, {Doel}, {Drlica-Wagner}, {Eifler}, {Evrard}, {Fernandez},
  {Filippenko}, {Finley}, {Flaugher}, {Foley}, {Fosalba}, {Frieman}, {Galbany},
  {Garcia-Bellido}, {Gaztanaga}, {Giannantonio}, {Glazebrook}, {Goldstein},
  {Gonzalez-Gaitan}, {Gruen}, {Gruendl}, {Gschwend}, {Gupta}, {Gutierrez},
  {Hartley}, {Hinton}, {Hollowood}, {Honscheid}, {Hoormann}, {Hoyle}, {James},
  {Jeltema}, {Johnson}, {Johnson}, {Kasai}, {Kent}, {Kessler}, {Kim},
  {Kirshner}, {Kovacs}, {Krause}, {Kron}, {Kuehn}, {Kuhlmann}, {Kuropatkin},
  {Lahav}, {Lasker}, {Lewis}, {Li}, {Lidman}, {Lima}, {Lin}, {Macaulay},
  {Maia}, {Mandel}, {March}, {Marriner}, {Marshall}, {Martini}, {Menanteau},
  {Miller}, {Miquel}, {Miranda}, {Mohr}, {Morganson}, {Muthukrishna},
  {M{\"o}ller}, {Neilsen}, {Nichol}, {Nord}, {Nugent}, {Ogando}, {Palmese},
  {Pan}, {Plazas}, {Pursiainen}, {Romer}, {Roodman}, {Rozo}, {Rykoff}, {Sako},
  {Sanchez}, {Scarpine}, {Schindler}, {Schubnell}, {Scolnic}, {Serrano},
  {Sevilla-Noarbe}, {Sharp}, {Smith}, {Soares-Santos}, {Sobreira}, {Sommer},
  {Spinka}, {Suchyta}, {Sullivan}, {Swann}, {Tarle}, {Thomas}, {Thomas},
  {Troxel}, {Tucker}, {Uddin}, {Walker}, {Wiseman}, {Wolf}, {Yanny}, {Zhang},
  \& {Zhang}}]{DES18}
{DES Collaboration}, {Abbott}, T.~M.~C., {Allam}, S., {et~al.} 2018, ArXiv
  e-prints, arXiv:1811.02374

\bibitem[{{Dilday} {et~al.}(2012){Dilday}, {Howell}, {Cenko}, {Silverman},
  {Nugent}, {Sullivan}, {Ben-Ami}, {Bildsten}, {Bolte}, {Endl}, {Filippenko},
  {Gnat}, {Horesh}, {Hsiao}, {Kasliwal}, {Kirkman}, {Maguire}, {Marcy},
  {Moore}, {Pan}, {Parrent}, {Podsiadlowski}, {Quimby}, {Sternberg}, {Suzuki},
  {Tytler}, {Xu}, {Bloom}, {Gal-Yam}, {Hook}, {Kulkarni}, {Law}, {Ofek},
  {Polishook}, \& {Poznanski}}]{Dilday12Sci}
{Dilday}, B., {Howell}, D.~A., {Cenko}, S.~B., {et~al.} 2012, Science, 337, 942

\bibitem[{{Dimitriadis} {et~al.}(2018){Dimitriadis}, {Foley}, {Rest}, {Kasen},
  {Piro}, {Polin}, {Jones}, {Villar}, {Narayan}, {Coulter}, {Kilpatrick},
  {Pan}, {Rojas-Bravo}, {Fox}, {Jha}, {Nugent}, {Riess}, {Scolnic}, {Drout},
  {Barentsen}, {Dotson}, {Gully-Santiago}, {Hedges}, {Cody}, {Barclay},
  {Howell}, {Garnavich}, {Tucker}, {Shaya}, {Mushotzky}, {Olling}, {Margheim},
  {Zenteno}, {Coughlin}, {Van Cleve}, {Cardoso}, {Larson}, {McCalmont-Everton},
  {Peterson}, {Ross}, {Reedy}, {Osborne}, {McGinn}, {Kohnert}, {Migliorini},
  {Wheaton}, {Spencer}, {Labonde}, {Castillo}, {Beerman}, {Steward}, {Hanley},
  {Larsen}, {Gangopadhyay}, {Kloetzel}, {Weschler}, {Nystrom}, {Moffatt},
  {Redick}, {Griest}, {Packard}, {Muszynski}, {Kampmeier}, {Bjella}, {Flynn},
  {Elsaesser}, {Chambers}, {Flewelling}, {Huber}, {Magnier}, {Waters},
  {Schultz}, {Bulger}, {Lowe}, {Willman}, {Smartt}, {Smith}, {Points},
  {Strampelli}, {Brimacombe}, {Chen}, {Munoz}, {Mutel}, {Shields}, {Vallely},
  {Villanueva}, {Li}, {Wang}, {Zhang}, {Lin}, {Mo}, {Zhao}, {Sai}, {Zhang},
  {Zhang}, {Zhang}, {Wang}, {Zhang}, {Baron}, {DerKacy}, {Li}, {Chen}, {Xiang},
  {Rui}, {Wang}, {Huang}, {Li}, {Hosseinzadeh}, {Howell}, {Arcavi},
  {Hiramatsu}, {Burke}, {Valenti}, {Tonry}, {Denneau}, {Heinze}, {Weiland},
  {Stalder}, {Vinko}, {Sarneczky}, {Pa}, {Bodi}, {Bognar}, {Csak}, {Cseh},
  {Csornyei}, {Hanyecz}, {Ignacz}, {Kalup}, {Konyves-Toth}, {Kriskovics},
  {Ordasi}, {Rajmon}, {Sodor}, {Szabo}, {Szakats}, {Zsidi}, {Williams},
  {Nordin}, {Cartier}, {Frohmaier}, {Galbany}, {Gutierrez}, {Hook}, {Inserra},
  {Smith}, {Sand}, {Andrews}, {Smith}, \& {Bilinski}}]{Dimitriadis18arXiv}
{Dimitriadis}, G., {Foley}, R.~J., {Rest}, A., {et~al.} 2018, ArXiv e-prints,
  arXiv:1811.10061

\bibitem[{{Dotson} {et~al.}(2018){Dotson}, {Rest}, {Barentsen},
  {Gully-Santiago}, {Fleming}, {Garnavich}, {Tucker}, {Kasen}, {Narayan},
  {Shaya}, {Olling}, {Margheim}, {Zenteno}, {Villar}, {Chambers}, {Flewelling},
  {Huber}, {Magnier}, {Waters}, {Schultz}, {Bulger}, {Lowe}, {Willman},
  {Smartt}, \& {Smith}}]{Dotson18RNAAS}
{Dotson}, J.~L., {Rest}, A., {Barentsen}, G., {et~al.} 2018, Research Notes of
  the American Astronomical Society, 2, 178

\bibitem[{{Faber} {et~al.}(2003){Faber}, {Phillips}, {Kibrick}, {Alcott},
  {Allen}, {Burrous}, {Cantrall}, {Clarke}, {Coil}, {Cowley}, {Davis}, {Deich},
  {Dietsch}, {Gilmore}, {Harper}, {Hilyard}, {Lewis}, {McVeigh}, {Newman},
  {Osborne}, {Schiavon}, {Stover}, {Tucker}, {Wallace}, {Wei}, {Wirth}, \&
  {Wright}}]{Faber03SPIE}
{Faber}, S.~M., {Phillips}, A.~C., {Kibrick}, R.~I., {et~al.} 2003, in
  \procspie, Vol. 4841, Instrument Design and Performance for Optical/Infrared
  Ground-based Telescopes, ed. M.~{Iye} \& A.~F.~M. {Moorwood}, 1657--1669

\bibitem[{{Firth} {et~al.}(2015){Firth}, {Sullivan}, {Gal-Yam}, {Howell},
  {Maguire}, {Nugent}, {Piro}, {Baltay}, {Feindt}, {Hadjiyksta}, {McKinnon},
  {Ofek}, {Rabinowitz}, \& {Walker}}]{Firth15MNRAS}
{Firth}, R.~E., {Sullivan}, M., {Gal-Yam}, A., {et~al.} 2015, \mnras, 446, 3895

\bibitem[{{Fitzpatrick}(1999)}]{Fitzpatrick99PASP}
{Fitzpatrick}, E.~L. 1999, \pasp, 111, 63

\bibitem[{{Foley} {et~al.}(2016){Foley}, {Jha}, {Pan}, {Zheng}, {Bildsten},
  {Filippenko}, \& {Kasen}}]{Foley16MNRAS}
{Foley}, R.~J., {Jha}, S.~W., {Pan}, Y.-C., {et~al.} 2016, \mnras, 461, 433

\bibitem[{{Foley} {et~al.}(2003){Foley}, {Papenkova}, {Swift}, {Filippenko},
  {Li}, {Mazzali}, {Chornock}, {Leonard}, \& {Van Dyk}}]{Foley03PASP}
{Foley}, R.~J., {Papenkova}, M.~S., {Swift}, B.~J., {et~al.} 2003, \pasp, 115,
  1220

\bibitem[{{Foley} {et~al.}(2012{\natexlab{a}}){Foley}, {Challis}, {Filippenko},
  {Ganeshalingam}, {Landsman}, {Li}, {Marion}, {Silverman}, {Beaton},
  {Bennert}, {Cenko}, {Childress}, {Guhathakurta}, {Jiang}, {Kalirai},
  {Kirshner}, {Stockton}, {Tollerud}, {Vink{\'o}}, {Wheeler}, \&
  {Woo}}]{Foley12ApJ}
{Foley}, R.~J., {Challis}, P.~J., {Filippenko}, A.~V., {et~al.}
  2012{\natexlab{a}}, \apj, 744, 38

\bibitem[{{Foley} {et~al.}(2012{\natexlab{b}}){Foley}, {Simon}, {Burns},
  {Gal-Yam}, {Hamuy}, {Kirshner}, {Morrell}, {Phillips}, {Shields}, \&
  {Sternberg}}]{Foley12ApJ2}
{Foley}, R.~J., {Simon}, J.~D., {Burns}, C.~R., {et~al.} 2012{\natexlab{b}},
  \apj, 752, 101

\bibitem[{{Foley} {et~al.}(2013){Foley}, {Challis}, {Chornock},
  {Ganeshalingam}, {Li}, {Marion}, {Morrell}, {Pignata}, {Stritzinger},
  {Silverman}, {Wang}, {Anderson}, {Filippenko}, {Freedman}, {Hamuy}, {Jha},
  {Kirshner}, {McCully}, {Persson}, {Phillips}, {Reichart}, \&
  {Soderberg}}]{Foley13ApJ}
{Foley}, R.~J., {Challis}, P.~J., {Chornock}, R., {et~al.} 2013, \apj, 767, 57

\bibitem[{{Ganeshalingam} {et~al.}(2011){Ganeshalingam}, {Li}, \&
  {Filippenko}}]{Ganeshalingam11MNRAS}
{Ganeshalingam}, M., {Li}, W., \& {Filippenko}, A.~V. 2011, \mnras, 416, 2607

\bibitem[{{Gonz{\'a}lez-Gait{\'a}n} {et~al.}(2012){Gonz{\'a}lez-Gait{\'a}n},
  {Conley}, {Bianco}, {Howell}, {Sullivan}, {Perrett}, {Carlberg}, {Astier},
  {Balam}, {Balland}, {Basa}, {Fouchez}, {Fourmanoit}, {Graham}, {Guy},
  {Hardin}, {Hook}, {Lidman}, {Pain}, {Palanque-Delabrouille}, {Pritchet},
  {Regnault}, {Rich}, \& {Ruhlmann-Kleider}}]{GonzalezGaitan12}
{Gonz{\'a}lez-Gait{\'a}n}, S., {Conley}, A., {Bianco}, F.~B., {et~al.} 2012,
  \apj, 745, 44

\bibitem[{{Graham} {et~al.}(2015){Graham}, {Nugent}, {Sullivan}, {Filippenko},
  {Cenko}, {Silverman}, {Clubb}, \& {Zheng}}]{Graham15MNRAS}
{Graham}, M.~L., {Nugent}, P.~E., {Sullivan}, M., {et~al.} 2015, \mnras, 454,
  1948

\bibitem[{{Graham} {et~al.}(2017){Graham}, {Kumar}, {Hosseinzadeh},
  {Hiramatsu}, {Arcavi}, {Howell}, {Valenti}, {Sand}, {Parrent}, {McCully}, \&
  {Filippenko}}]{Graham17MNRAS}
{Graham}, M.~L., {Kumar}, S., {Hosseinzadeh}, G., {et~al.} 2017, \mnras, 472,
  3437

\bibitem[{{Haas} {et~al.}(2010){Haas}, {Batalha}, {Bryson}, {Caldwell},
  {Dotson}, {Hall}, {Jenkins}, {Klaus}, {Koch}, {Kolodziejczak}, {Middour},
  {Smith}, {Sobeck}, {Stober}, {Thompson}, \& {Van Cleve}}]{Haas10ApJ}
{Haas}, M.~R., {Batalha}, N.~M., {Bryson}, S.~T., {et~al.} 2010, \apjl, 713,
  L115

\bibitem[{{Hayden} {et~al.}(2010){Hayden}, {Garnavich}, {Kessler}, {Frieman},
  {Jha}, {Bassett}, {Cinabro}, {Dilday}, {Kasen}, {Marriner}, {Nichol},
  {Riess}, {Sako}, {Schneider}, {Smith}, \& {Sollerman}}]{Hayden10ApJ1}
{Hayden}, B.~T., {Garnavich}, P.~M., {Kessler}, R., {et~al.} 2010, \apj, 712,
  350

\bibitem[{{Hillebrandt} {et~al.}(2013){Hillebrandt}, {Kromer}, {R{\"o}pke}, \&
  {Ruiter}}]{Hillebrandt13FrPhy}
{Hillebrandt}, W., {Kromer}, M., {R{\"o}pke}, F.~K., \& {Ruiter}, A.~J. 2013,
  Frontiers of Physics, 8, 116

\bibitem[{{Hosseinzadeh} {et~al.}(2017){Hosseinzadeh}, {Sand}, {Valenti},
  {Brown}, {Howell}, {McCully}, {Kasen}, {Arcavi}, {Azalee Bostroem},
  {Tartaglia}, {Hsiao}, {Davis}, {Shahbandeh}, \&
  {Stritzinger}}]{Hosseinzadeh17ApJ}
{Hosseinzadeh}, G., {Sand}, D.~J., {Valenti}, S., {et~al.} 2017, \apjl, 845,
  L11

\bibitem[{{Hoyle} \& {Fowler}(1960)}]{Hoyle60}
{Hoyle}, F., \& {Fowler}, W.~A. 1960, \apj, 132, 565

\bibitem[{{Iben} \& {Tutukov}(1984)}]{Iben84ApJS}
{Iben}, Jr., I., \& {Tutukov}, A.~V. 1984, \apjs, 54, 335

\bibitem[{{Jiang} {et~al.}(2017){Jiang}, {Doi}, {Maeda}, {Shigeyama}, {Nomoto},
  {Yasuda}, {Jha}, {Tanaka}, {Morokuma}, {Tominaga}, {Ivezi{\'c}},
  {Ruiz-Lapuente}, {Stritzinger}, {Mazzali}, {Ashall}, {Mould}, {Baade},
  {Suzuki}, {Connolly}, {Patat}, {Wang}, {Yoachim}, {Jones}, {Furusawa}, \&
  {Miyazaki}}]{Jiang17Natur}
{Jiang}, J.-A., {Doi}, M., {Maeda}, K., {et~al.} 2017, \nat, 550, 80

\bibitem[{{Jones} {et~al.}(2018){Jones}, {Scolnic}, {Riess}, {Rest},
  {Kirshner}, {Berger}, {Kessler}, {Pan}, {Foley}, {Chornock}, {Ortega},
  {Challis}, {Burgett}, {Chambers}, {Draper}, {Flewelling}, {Huber}, {Kaiser},
  {Kudritzki}, {Metcalfe}, {Tonry}, {Wainscoat}, {Waters}, {Gall}, {Kotak},
  {McCrum}, {Smartt}, \& {Smith}}]{Jones18}
{Jones}, D.~O., {Scolnic}, D.~M., {Riess}, A.~G., {et~al.} 2018, \apj, 857, 51

\bibitem[{{Kasen}(2010)}]{Kasen10ApJ}
{Kasen}, D. 2010, \apj, 708, 1025

\bibitem[{{Kasen} {et~al.}(2009){Kasen}, {R{\"o}pke}, \&
  {Woosley}}]{Kasen09Natur}
{Kasen}, D., {R{\"o}pke}, F.~K., \& {Woosley}, S.~E. 2009, \nat, 460, 869

\bibitem[{{Leadbeater}(2018)}]{Leadbeater18TNS}
{Leadbeater}, R. 2018, Transient Name Server Classification Report, 159

\bibitem[{{Leonard}(2007)}]{Leonard07ApJ}
{Leonard}, D.~C. 2007, \apj, 670, 1275

\bibitem[{{Levanon} \& {Soker}(2017)}]{Levanon17MNRAS}
{Levanon}, N., \& {Soker}, N. 2017, \mnras, 470, 2510

\bibitem[{{Li} {et~al.}(2018){Li}, {Wang}, {Vink{\'o}}, {Mo}, {Hosseinzadeh},
  {Sand}, {Zhang}, {Lin}, {Zhang}, {Wang}, {Zhang}, {Chen}, {Xiang}, {Rui},
  {Huang}, {Li}, {Zhang}, {Li}, {Baron}, {Derkacy}, {Zhao}, {Sai}, {Zhang},
  {Wang}, {Howell}, {McCully}, {Arcavi}, {Valenti}, {Hiramatsu}, {Burke},
  {Rest}, {Garnavich}, {Tucker}, {Narayan}, {Shaya}, {Margheim}, {Zenteno},
  {Villar}, {Dimitriadis}, {Foley}, {Pan}, {Coulter}, {Fox}, {Jha}, {Jones},
  {Kasen}, {Kilpatrick}, {Piro}, {Riess}, {Rojas-Bravo}, {Shappee}, {Holoien},
  {Stanek}, {Drout}, {Auchettl}, {Kochanek}, {Brown}, {Bose}, {Bersier},
  {Brimacombe}, {Chen}, {Dong}, {Holmbo}, {Mu{\~n}oz}, {Mutel}, {Post},
  {Prieto}, {Shields}, {Tallon}, {Thompson}, {Vallely}, {Villanueva}, {Smartt},
  {Smith}, {Chambers}, {Flewelling}, {Huber}, {Magnier}, {Waters}, {Schultz},
  {Bulger}, {Lowe}, {Willman}, {S{\'a}rneczky}, {P{\'a}l}, {Wheeler},
  {B{\'o}di}, {Bogn{\'a}r}, {Cs{\'a}k}, {Cseh}, {Cs{\"o}rnyei}, {Hanyecz},
  {Ign{\'a}cz}, {Kalup}, {K{\"o}nyves-T{\'o}th}, {Kriskovics}, {Ordasi},
  {Rajmon}, {S{\'o}dor}, {Szab{\'o}}, {Szak{\'a}ts}, {Zsidi}, {Milne},
  {Andrews}, {Smith}, {Bilinski}, {Brown}, {Nordin}, {Williams}, {Galbany},
  {Palmerio}, {Hook}, {Inserra}, {Maguire}, {Cartier}, {Razza},
  {Guti{\'e}rrez}, {Hermes}, {Reding}, {Kaiser}, {Tonry}, {Heinze}, {Denneau},
  {Weiland}, {Stalder}, {Barentsen}, {Dotson}, {Barclay}, {Gully-Santiago},
  {Hedges}, {Cody}, {Howell}, {Coughlin}, {Van Cleve}, {Cardoso}, {Larson},
  {McCalmont-Everton}, {Peterson}, {Ross}, {Reedy}, {Osborne}, {McGinn},
  {Kohnert}, {Migliorini}, {Wheaton}, {Spencer}, {Labonde}, {Castillo},
  {Beerman}, {Steward}, {Hanley}, {Larsen}, {Gangopadhyay}, {Kloetzel},
  {Weschler}, {Nystrom}, {Moffatt}, {Redick}, {Griest}, {Packard}, {Muszynski},
  {Kampmeier}, {Bjella}, {Flynn}, \& {Elsaesser}}]{Li18arXiv}
{Li}, W., {Wang}, X., {Vink{\'o}}, J., {et~al.} 2018, ArXiv e-prints,
  arXiv:1811.10056

\bibitem[{{Liu} {et~al.}(2013){Liu}, {Pakmor}, {Seitenzahl}, {Hillebrandt},
  {Kromer}, {R{\"o}pke}, {Edelmann}, {Taubenberger}, {Maeda}, {Wang}, \&
  {Han}}]{Liu13ApJ}
{Liu}, Z.-W., {Pakmor}, R., {Seitenzahl}, I.~R., {et~al.} 2013, \apj, 774, 37

\bibitem[{{Lundqvist} {et~al.}(2013){Lundqvist}, {Mattila}, {Sollerman},
  {Kozma}, {Baron}, {Cox}, {Fransson}, {Leibundgut}, \&
  {Spyromilio}}]{Lundqvist13MNRAS}
{Lundqvist}, P., {Mattila}, S., {Sollerman}, J., {et~al.} 2013, \mnras, 435,
  329

\bibitem[{{Magnier} {et~al.}(2016){Magnier}, {Chambers}, {Flewelling},
  {Hoblitt}, {Huber}, {Price}, {Sweeney}, {Waters}, {Denneau}, {Draper},
  {Hodapp}, {Jedicke}, {Kaiser}, {Kudritzki}, {Metcalfe}, {Stubbs}, \&
  {Wainscoast}}]{Magnier16arXiv}
{Magnier}, E.~A., {Chambers}, K.~C., {Flewelling}, H.~A., {et~al.} 2016, ArXiv
  e-prints, arXiv:1612.05240

\bibitem[{{Maguire} {et~al.}(2016){Maguire}, {Taubenberger}, {Sullivan}, \&
  {Mazzali}}]{Maguire16MNRAS}
{Maguire}, K., {Taubenberger}, S., {Sullivan}, M., \& {Mazzali}, P.~A. 2016,
  \mnras, 457, 3254

\bibitem[{{Maguire} {et~al.}(2013){Maguire}, {Sullivan}, {Patat}, {Gal-Yam},
  {Hook}, {Dhawan}, {Howell}, {Mazzali}, {Nugent}, {Pan}, {Podsiadlowski},
  {Simon}, {Sternberg}, {Valenti}, {Baltay}, {Bersier}, {Blagorodnova}, {Chen},
  {Ellman}, {Feindt}, {F{\"o}rster}, {Fraser}, {Gonz{\'a}lez-Gait{\'a}n},
  {Graham}, {Guti{\'e}rrez}, {Hachinger}, {Hadjiyska}, {Inserra}, {Knapic},
  {Laher}, {Leloudas}, {Margheim}, {McKinnon}, {Molinaro}, {Morrell}, {Ofek},
  {Rabinowitz}, {Rest}, {Sand}, {Smareglia}, {Smartt}, {Taddia}, {Walker},
  {Walton}, \& {Young}}]{Maguire13MNRAS}
{Maguire}, K., {Sullivan}, M., {Patat}, F., {et~al.} 2013, \mnras, 436, 222

\bibitem[{{Marietta} {et~al.}(2000){Marietta}, {Burrows}, \&
  {Fryxell}}]{Marietta00ApJS}
{Marietta}, E., {Burrows}, A., \& {Fryxell}, B. 2000, \apjs, 128, 615

\bibitem[{{Marion} {et~al.}(2016){Marion}, {Brown}, {Vink{\'o}}, {Silverman},
  {Sand}, {Challis}, {Kirshner}, {Wheeler}, {Berlind}, {Brown}, {Calkins},
  {Camacho}, {Dhungana}, {Foley}, {Friedman}, {Graham}, {Howell}, {Hsiao},
  {Irwin}, {Jha}, {Kehoe}, {Macri}, {Maeda}, {Mandel}, {McCully}, {Pandya},
  {Rines}, {Wilhelmy}, \& {Zheng}}]{Marion16ApJ}
{Marion}, G.~H., {Brown}, P.~J., {Vink{\'o}}, J., {et~al.} 2016, \apj, 820, 92

\bibitem[{{Mattila} {et~al.}(2005){Mattila}, {Lundqvist}, {Sollerman}, {Kozma},
  {Baron}, {Fransson}, {Leibundgut}, \& {Nomoto}}]{Mattila05AA}
{Mattila}, S., {Lundqvist}, P., {Sollerman}, J., {et~al.} 2005, \aap, 443, 649

\bibitem[{{McCully} {et~al.}(2014){McCully}, {Jha}, {Foley}, {Bildsten},
  {Fong}, {Kirshner}, {Marion}, {Riess}, \& {Stritzinger}}]{McCully14Natur}
{McCully}, C., {Jha}, S.~W., {Foley}, R.~J., {et~al.} 2014, \nat, 512, 54

\bibitem[{{Meng} \& {Li}(2018)}]{Meng18arXiv}
{Meng}, X., \& {Li}, J. 2018, arXiv e-prints, arXiv:1811.11351

\bibitem[{{Meng} \& {Podsiadlowski}(2018)}]{Meng18ApJ}
{Meng}, X., \& {Podsiadlowski}, P. 2018, \apj, 861, 127

\bibitem[{{Miller} {et~al.}(2018){Miller}, {Cao}, {Piro}, {Blagorodnova},
  {Bue}, {Cenko}, {Dhawan}, {Ferretti}, {Fox}, {Fremling}, {Goobar}, {Howell},
  {Hosseinzadeh}, {Kasliwal}, {Laher}, {Lunnan}, {Masci}, {McCully}, {Nugent},
  {Sollerman}, {Taddia}, \& {Kulkarni}}]{Miller18ApJ}
{Miller}, A.~A., {Cao}, Y., {Piro}, A.~L., {et~al.} 2018, \apj, 852, 100

\bibitem[{{Munari} {et~al.}(2013){Munari}, {Henden}, {Belligoli}, {Castellani},
  {Cherini}, {Righetti}, \& {Vagnozzi}}]{Munari13NewA}
{Munari}, U., {Henden}, A., {Belligoli}, R., {et~al.} 2013, \na, 20, 30

\bibitem[{{Newman} {et~al.}(2013){Newman}, {Cooper}, {Davis}, {Faber}, {Coil},
  {Guhathakurta}, {Koo}, {Phillips}, {Conroy}, {Dutton}, {Finkbeiner}, {Gerke},
  {Rosario}, {Weiner}, {Willmer}, {Yan}, {Harker}, {Kassin}, {Konidaris},
  {Lai}, {Madgwick}, {Noeske}, {Wirth}, {Connolly}, {Kaiser}, {Kirby},
  {Lemaux}, {Lin}, {Lotz}, {Luppino}, {Marinoni}, {Matthews}, {Metevier}, \&
  {Schiavon}}]{Newman13ApJS}
{Newman}, J.~A., {Cooper}, M.~C., {Davis}, M., {et~al.} 2013, \apjs, 208, 5

\bibitem[{{Nugent} {et~al.}(2011){Nugent}, {Sullivan}, {Cenko}, {Thomas},
  {Kasen}, {Howell}, {Bersier}, {Bloom}, {Kulkarni}, {Kandrashoff},
  {Filippenko}, {Silverman}, {Marcy}, {Howard}, {Isaacson}, {Maguire},
  {Suzuki}, {Tarlton}, {Pan}, {Bildsten}, {Fulton}, {Parrent}, {Sand},
  {Podsiadlowski}, {Bianco}, {Dilday}, {Graham}, {Lyman}, {James}, {Kasliwal},
  {Law}, {Quimby}, {Hook}, {Walker}, {Mazzali}, {Pian}, {Ofek}, {Gal-Yam}, \&
  {Poznanski}}]{Nugent11}
{Nugent}, P.~E., {Sullivan}, M., {Cenko}, S.~B., {et~al.} 2011, \nat, 480, 344

\bibitem[{{Oke} {et~al.}(1995){Oke}, {Cohen}, {Carr}, {Cromer}, {Dingizian},
  {Harris}, {Labrecque}, {Lucinio}, {Schaal}, {Epps}, \& {Miller}}]{Oke95PASP}
{Oke}, J.~B., {Cohen}, J.~G., {Carr}, M., {et~al.} 1995, \pasp, 107, 375

\bibitem[{{Olling} {et~al.}(2015){Olling}, {Mushotzky}, {Shaya}, {Rest},
  {Garnavich}, {Tucker}, {Kasen}, {Margheim}, \& {Filippenko}}]{Olling15Natur}
{Olling}, R.~P., {Mushotzky}, R., {Shaya}, E.~J., {et~al.} 2015, \nat, 521, 332

\bibitem[{{Pan} {et~al.}(2012){Pan}, {Ricker}, \& {Taam}}]{Pan12ApJ}
{Pan}, K.-C., {Ricker}, P.~M., \& {Taam}, R.~E. 2012, \apj, 750, 151

\bibitem[{{Patat} {et~al.}(2007){Patat}, {Chandra}, {Chevalier}, {Justham},
  {Podsiadlowski}, {Wolf}, {Gal-Yam}, {Pasquini}, {Crawford}, {Mazzali},
  {Pauldrach}, {Nomoto}, {Benetti}, {Cappellaro}, {Elias-Rosa}, {Hillebrandt},
  {Leonard}, {Pastorello}, {Renzini}, {Sabbadin}, {Simon}, \&
  {Turatto}}]{Patat07Sci}
{Patat}, F., {Chandra}, P., {Chevalier}, R., {et~al.} 2007, Science, 317, 924

\bibitem[{{Rest} {et~al.}(2005){Rest}, {Stubbs}, {Becker}, {Miknaitis},
  {Miceli}, {Covarrubias}, {Hawley}, {Smith}, {Suntzeff}, {Olsen}, {Prieto},
  {Hiriart}, {Welch}, {Cook}, {Nikolaev}, {Huber}, {Prochtor}, {Clocchiatti},
  {Minniti}, {Garg}, {Challis}, {Keller}, \& {Schmidt}}]{Rest05ApJ}
{Rest}, A., {Stubbs}, C., {Becker}, A.~C., {et~al.} 2005, \apj, 634, 1103

\bibitem[{{Rest} {et~al.}(2014){Rest}, {Scolnic}, {Foley}, {Huber}, {Chornock},
  {Narayan}, {Tonry}, {Berger}, {Soderberg}, {Stubbs}, {Riess}, {Kirshner},
  {Smartt}, {Schlafly}, {Rodney}, {Botticella}, {Brout}, {Challis}, {Czekala},
  {Drout}, {Hudson}, {Kotak}, {Leibler}, {Lunnan}, {Marion}, {McCrum},
  {Milisavljevic}, {Pastorello}, {Sanders}, {Smith}, {Stafford}, {Thilker},
  {Valenti}, {Wood-Vasey}, {Zheng}, {Burgett}, {Chambers}, {Denneau}, {Draper},
  {Flewelling}, {Hodapp}, {Kaiser}, {Kudritzki}, {Magnier}, {Metcalfe},
  {Price}, {Sweeney}, {Wainscoat}, \& {Waters}}]{Rest14ApJ}
{Rest}, A., {Scolnic}, D., {Foley}, R.~J., {et~al.} 2014, \apj, 795, 44

\bibitem[{{Riess} {et~al.}(1999){Riess}, {Filippenko}, {Li}, {Treffers},
  {Schmidt}, {Qiu}, {Hu}, {Armstrong}, {Faranda}, {Thouvenot}, \&
  {Buil}}]{Riess99AJ}
{Riess}, A.~G., {Filippenko}, A.~V., {Li}, W., {et~al.} 1999, \aj, 118, 2675

\bibitem[{{Riess} {et~al.}(2016){Riess}, {Macri}, {Hoffmann}, {Scolnic},
  {Casertano}, {Filippenko}, {Tucker}, {Reid}, {Jones}, {Silverman},
  {Chornock}, {Challis}, {Yuan}, {Brown}, \& {Foley}}]{Riess16ApJ}
{Riess}, A.~G., {Macri}, L.~M., {Hoffmann}, S.~L., {et~al.} 2016, \apj, 826, 56

\bibitem[{{Sand} {et~al.}(2018){Sand}, {Graham}, {Boty{\'a}nszki}, {Hiramatsu},
  {McCully}, {Valenti}, {Hosseinzadeh}, {Howell}, {Burke}, {Cartier},
  {Diamond}, {Hsiao}, {Jha}, {Kasen}, {Kumar}, {Marion}, {Suntzeff},
  {Tartaglia}, {Wheeler}, \& {Wyatt}}]{Sand18ApJ}
{Sand}, D.~J., {Graham}, M.~L., {Boty{\'a}nszki}, J., {et~al.} 2018, \apj, 863,
  24

\bibitem[{{Schechter} {et~al.}(1993){Schechter}, {Mateo}, \&
  {Saha}}]{Schechter93PASP}
{Schechter}, P.~L., {Mateo}, M., \& {Saha}, A. 1993, \pasp, 105, 1342

\bibitem[{{Scolnic} {et~al.}(2018){Scolnic}, {Jones}, {Rest}, {Pan},
  {Chornock}, {Foley}, {Huber}, {Kessler}, {Narayan}, {Riess}, {Rodney},
  {Berger}, {Brout}, {Challis}, {Drout}, {Finkbeiner}, {Lunnan}, {Kirshner},
  {Sanders}, {Schlafly}, {Smartt}, {Stubbs}, {Tonry}, {Wood-Vasey}, {Foley},
  {Hand}, {Johnson}, {Burgett}, {Chambers}, {Draper}, {Hodapp}, {Kaiser},
  {Kudritzki}, {Magnier}, {Metcalfe}, {Bresolin}, {Gall}, {Kotak}, {McCrum}, \&
  {Smith}}]{Scolnic18ApJ}
{Scolnic}, D.~M., {Jones}, D.~O., {Rest}, A., {et~al.} 2018, \apj, 859, 101

\bibitem[{{Shappee} {et~al.}(2018{\natexlab{a}}){Shappee}, {Piro}, {Stanek},
  {Patel}, {Margutti}, {Lipunov}, \& {Pogge}}]{Shappee18ApJ}
{Shappee}, B.~J., {Piro}, A.~L., {Stanek}, K.~Z., {et~al.} 2018{\natexlab{a}},
  \apj, 855, 6

\bibitem[{{Shappee} {et~al.}(2013){Shappee}, {Stanek}, {Pogge}, \&
  {Garnavich}}]{Shappee13ApJ}
{Shappee}, B.~J., {Stanek}, K.~Z., {Pogge}, R.~W., \& {Garnavich}, P.~M. 2013,
  \apjl, 762, L5

\bibitem[{{Shappee} {et~al.}(2014){Shappee}, {Prieto}, {Grupe}, {Kochanek},
  {Stanek}, {De Rosa}, {Mathur}, {Zu}, {Peterson}, {Pogge}, {Komossa}, {Im},
  {Jencson}, {Holoien}, {Basu}, {Beacom}, {Szczygie{\l}}, {Brimacombe},
  {Adams}, {Campillay}, {Choi}, {Contreras}, {Dietrich}, {Dubberley},
  {Elphick}, {Foale}, {Giustini}, {Gonzalez}, {Hawkins}, {Howell}, {Hsiao},
  {Koss}, {Leighly}, {Morrell}, {Mudd}, {Mullins}, {Nugent}, {Parrent},
  {Phillips}, {Pojmanski}, {Rosing}, {Ross}, {Sand}, {Terndrup}, {Valenti},
  {Walker}, \& {Yoon}}]{Shappee14ApJ}
{Shappee}, B.~J., {Prieto}, J.~L., {Grupe}, D., {et~al.} 2014, \apj, 788, 48

\bibitem[{{Shappee} {et~al.}(2016){Shappee}, {Piro}, {Holoien}, {Prieto},
  {Contreras}, {Itagaki}, {Burns}, {Kochanek}, {Stanek}, {Alper}, {Basu},
  {Beacom}, {Bersier}, {Brimacombe}, {Conseil}, {Danilet}, {Dong}, {Falco},
  {Grupe}, {Hsiao}, {Kiyota}, {Morrell}, {Nicolas}, {Phillips}, {Pojmanski},
  {Simonian}, {Stritzinger}, {Szczygie{\l}}, {Taddia}, {Thompson},
  {Thorstensen}, {Wagner}, \& {Wo{\'z}niak}}]{Shappee16ApJ}
{Shappee}, B.~J., {Piro}, A.~L., {Holoien}, T.~W.-S., {et~al.} 2016, \apj, 826,
  144

\bibitem[{{Shappee} {et~al.}(2018{\natexlab{b}}){Shappee}, {Holoien}, {Drout},
  {Auchettl}, {Stritzinger}, {Kochanek}, {Stanek}, {Shaya}, {Narayan}, {Brown},
  {Bose}, {Bersier}, {Brimacombe}, {Chen}, {Dong}, {Holmbo}, {Katz}, {Munnoz},
  {Mutel}, {Post}, {Prieto}, {Shields}, {Tallon}, {Thompson}, {Vallely},
  {Villanueva}, {Denneau}, {Flewelling}, {Heinze}, {Smith}, {Stalder}, {Tonry},
  {Weiland}, {Barclay}, {Barentsen}, {Cody}, {Dotson}, {Foerster}, {Garnavich},
  {Gully-santiago}, {Hedges}, {Howell}, {Kasen}, {Margheim}, {Mushotzky},
  {Rest}, {Tucker}, {Villar}, {Zenteno}, {Beerman}, {Bjella}, {Castillo},
  {Coughlin}, {Elsaesser}, {Flynn}, {Gangopadhyay}, {Griest}, {Hanley},
  {Kampmeier}, {Kloetzel}, {Kohnert}, {Labonde}, {Larsen}, {Larson},
  {Mccalmont-everton}, {Mcginn}, {Migliorini}, {Moffatt}, {Muszynski},
  {Nystrom}, {Osborne}, {Packard}, {Peterson}, {Redick}, {Reedy}, {Ross},
  {Spencer}, {Steward}, {Van Cleve}, {Cardoso}, {Weschler}, {Wheaton},
  {Bulger}, {Lowe}, {Magnier}, {Schultz}, {Waters}, {Willman}, {Baron}, {Chen},
  {Derkacy}, {Huang}, {Li}, {Li}, {Li}, {Rui}, {Sai}, {Wang}, {Wang}, {Wang},
  {Xiang}, {Zhang}, {Zhang}, {Zhang}, {Zhang}, {Zhang}, {Zhao}, {Brown},
  {Hermes}, {Nordin}, {Points}, {Strampelli}, \& {Zenteno}}]{Shappee18arXiv}
{Shappee}, B.~J., {Holoien}, T.~W.-s., {Drout}, M.~R., {et~al.}
  2018{\natexlab{b}}, ArXiv e-prints, arXiv:1807.11526

\bibitem[{{Silverman} {et~al.}(2012){Silverman}, {Foley}, {Filippenko},
  {Ganeshalingam}, {Barth}, {Chornock}, {Griffith}, {Kong}, {Lee}, {Leonard},
  {Matheson}, {Miller}, {Steele}, {Barris}, {Bloom}, {Cobb}, {Coil},
  {Desroches}, {Gates}, {Ho}, {Jha}, {Kandrashoff}, {Li}, {Mandel}, {Modjaz},
  {Moore}, {Mostardi}, {Papenkova}, {Park}, {Perley}, {Poznanski}, {Reuter},
  {Scala}, {Serduke}, {Shields}, {Swift}, {Tonry}, {Van Dyk}, {Wang}, \&
  {Wong}}]{Silverman12MNRAS}
{Silverman}, J.~M., {Foley}, R.~J., {Filippenko}, A.~V., {et~al.} 2012, \mnras,
  425, 1789

\bibitem[{{Silverman} {et~al.}(2013){Silverman}, {Nugent}, {Gal-Yam},
  {Sullivan}, {Howell}, {Filippenko}, {Arcavi}, {Ben-Ami}, {Bloom}, {Cenko},
  {Cao}, {Chornock}, {Clubb}, {Coil}, {Foley}, {Graham}, {Griffith}, {Horesh},
  {Kasliwal}, {Kulkarni}, {Leonard}, {Li}, {Matheson}, {Miller}, {Modjaz},
  {Ofek}, {Pan}, {Perley}, {Poznanski}, {Quimby}, {Steele}, {Sternberg}, {Xu},
  \& {Yaron}}]{Silverman13ApJS}
{Silverman}, J.~M., {Nugent}, P.~E., {Gal-Yam}, A., {et~al.} 2013, \apjs, 207,
  3

\bibitem[{{Sim} {et~al.}(2013){Sim}, {Seitenzahl}, {Kromer},
  {Ciaraldi-Schoolmann}, {R{\"o}pke}, {Fink}, {Hillebrandt}, {Pakmor},
  {Ruiter}, \& {Taubenberger}}]{Sim13MNRAS}
{Sim}, S.~A., {Seitenzahl}, I.~R., {Kromer}, M., {et~al.} 2013, \mnras, 436,
  333

\bibitem[{{Simon} {et~al.}(2009){Simon}, {Gal-Yam}, {Gnat}, {Quimby},
  {Ganeshalingam}, {Silverman}, {Blondin}, {Li}, {Filippenko}, {Wheeler},
  {Kirshner}, {Patat}, {Nugent}, {Foley}, {Vogt}, {Butler}, {Peek},
  {Rosolowsky}, {Herczeg}, {Sauer}, \& {Mazzali}}]{Simon09ApJ}
{Simon}, J.~D., {Gal-Yam}, A., {Gnat}, O., {et~al.} 2009, \apj, 702, 1157

\bibitem[{{Sternberg} {et~al.}(2011){Sternberg}, {Gal-Yam}, {Simon}, {Leonard},
  {Quimby}, {Phillips}, {Morrell}, {Thompson}, {Ivans}, {Marshall},
  {Filippenko}, {Marcy}, {Bloom}, {Patat}, {Foley}, {Yong}, {Penprase},
  {Beeler}, {Allende Prieto}, \& {Stringfellow}}]{Sternberg11Sci}
{Sternberg}, A., {Gal-Yam}, A., {Simon}, J.~D., {et~al.} 2011, Science, 333,
  856

\bibitem[{{Stritzinger} {et~al.}(2018){Stritzinger}, {Shappee}, {Piro},
  {Ashall}, {Baron}, {Hoeflich}, {Holmbo}, {Holoien}, {Phillips}, {Burns},
  {Contreras}, {Morrell}, \& {Tucker}}]{Stritzinger18ApJ}
{Stritzinger}, M.~D., {Shappee}, B.~J., {Piro}, A.~L., {et~al.} 2018, \apjl,
  864, L35

\bibitem[{{Tucker} {et~al.}(2018){Tucker}, {Shappee}, \&
  {Wisniewski}}]{Tucker18arXiv}
{Tucker}, M.~A., {Shappee}, B.~J., \& {Wisniewski}, J.~P. 2018, arXiv e-prints,
  arXiv:1811.09635

\bibitem[{{Waters} {et~al.}(2016){Waters}, {Magnier}, {Price}, {Chambers},
  {Burgett}, {Draper}, {Flewelling}, {Hodapp}, {Huber}, {Jedicke}, {Kaiser},
  {Kudritzki}, {Lupton}, {Metcalfe}, {Rest}, {Sweeney}, {Tonry}, {Wainscoat},
  {Wood-Vasey}, \& {Builders}}]{Waters16arXiv}
{Waters}, C.~Z., {Magnier}, E.~A., {Price}, P.~A., {et~al.} 2016, ArXiv
  e-prints, arXiv:1612.05245

\bibitem[{{Whelan} \& {Iben}(1973)}]{Whelan73ApJ}
{Whelan}, J., \& {Iben}, Jr., I. 1973, \apj, 186, 1007

\bibitem[{{Woosley} \& {Kasen}(2011)}]{Woosley11ApJ}
{Woosley}, S.~E., \& {Kasen}, D. 2011, \apj, 734, 38

\bibitem[{{Woosley} {et~al.}(1986){Woosley}, {Taam}, \& {Weaver}}]{Woosley86}
{Woosley}, S.~E., {Taam}, R.~E., \& {Weaver}, T.~A. 1986, \apj, 301, 601

\bibitem[{{Zhang} {et~al.}(2018){Zhang}, {Xin}, {Li}, {Wang}, {Tan}, {Zhang},
  {Zhou}, {Mo}, {Rui}, \& {Xiang}}]{Zhang18ATel}
{Zhang}, J., {Xin}, Y., {Li}, W., {et~al.} 2018, The Astronomer's Telegram,
  11267

\bibitem[{{Zheng} {et~al.}(2013){Zheng}, {Silverman}, {Filippenko}, {Kasen},
  {Nugent}, {Graham}, {Wang}, {Valenti}, {Ciabattari}, {Kelly}, {Fox},
  {Shivvers}, {Clubb}, {Cenko}, {Balam}, {Howell}, {Hsiao}, {Li}, {Marion},
  {Sand}, {Vinko}, {Wheeler}, \& {Zhang}}]{Zheng13ApJ}
{Zheng}, W., {Silverman}, J.~M., {Filippenko}, A.~V., {et~al.} 2013, \apjl,
  778, L15

\end{thebibliography}

\end{document}